\documentclass[%
 reprint,
 amsmath,amssymb,
 aps,
 superscriptaddress,
]{revtex4-2}

\usepackage{graphicx}
\usepackage{amssymb, amsmath}
\usepackage{url}
\usepackage[hidelinks]{hyperref}
\usepackage{color}

\newcommand\PS[1]{{\color{black} #1}}
\newcommand\LW[1]{{\color{black} #1}}

\begin{document}

\title{The information content of brain states is explained by\\ structural constraints on state energetics}

\author{Leon Weninger}
\affiliation{Department of Bioengineering, School of Engineering \& Applied Sciences, University of Pennsylvania, Philadelphia, PA 19104, USA}
\affiliation{Institute of Imaging \& Computer Vision, RWTH Aachen University, 52072 Aachen, Germany}

\author{Pragya Srivastava}
\affiliation{Department of Bioengineering, School of Engineering \& Applied Sciences, University of Pennsylvania, Philadelphia, PA 19104, USA}

\author{Dale Zhou}
\affiliation{Department of Neuroscience, Perelman School of Medicine, University of Pennsylvania, Philadelphia, PA 19104, USA}
\affiliation{Department of Bioengineering, School of Engineering \& Applied Sciences, University of Pennsylvania, Philadelphia, PA 19104, USA}

\author{Jason Z. Kim}
\affiliation{Department of Bioengineering, School of Engineering \& Applied Sciences, University of Pennsylvania, Philadelphia, PA 19104, USA}

\author{Eli J. Cornblath}
\affiliation{Perelman School of Medicine, University of Pennsylvania, Philadelphia, PA 19104, USA}

\author{Maxwell A. Bertolero}
\affiliation{Department of Psychiatry, Perelman School of Medicine, University of Pennsylvania, Philadelphia, PA 19104, USA}

\author{Ute Habel}
\affiliation{Department of Psychiatry, Psychotherapy and Psychosomatics, Medical Faculty, RWTH Aachen University, Aachen, Germany}
\affiliation{Institute of Neuroscience and Medicine 10, Research Centre Jülich, Jülich, Germany}

\author{Dorit Merhof}
\affiliation{Institute of Imaging \& Computer Vision, RWTH Aachen University, 52072 Aachen, Germany}

\author{Dani S. Bassett}
\affiliation{Department of Bioengineering, School of Engineering \& Applied Sciences, University of Pennsylvania, Philadelphia, PA 19104, USA}
\affiliation{Department of Physics \& Astronomy, College of Arts and Sciences, University of Pennsylvania, Philadelphia, PA 19104, USA}
\affiliation{Department of Psychiatry, Perelman School of Medicine, University of Pennsylvania, Philadelphia, PA 19104, USA}
\affiliation{Department of Neurology, Perelman School of Medicine, University of Pennsylvania, Philadelphia, PA 19104, USA}
\affiliation{Department of Electrical \& Systems Engineering, School of Engineering and Applied Sciences, University of Pennsylvania, Philadelphia, PA 19104, USA}
\affiliation{Santa Fe Institute, Santa Fe, NM 87501, USA}
\affiliation{To whom correspondence should be addressed: dsb@seas.upenn.edu}

\date{\today}
\begin{abstract}

\noindent
Signal propagation along the structural connectome of the brain induces changes in the patterns of activity. 
These activity patterns define global brain states and contain information in accordance with their expected probability of occurrence. 
Being the physical substrate upon which information propagates, the structural connectome, in conjunction with the dynamics, determines the set of possible brain states and constrains the transition between accessible states.
Yet, precisely how these structural constraints on state-transitions relate to their information content remains unexplored.
To address this gap in knowledge, we defined the information content as a function of the activation distribution, where statistically rare values of activation correspond to high information content.
With this numerical definition in hand, we studied the spatiotemporal distribution of information content in fMRI data from the Human Connectome Project during different tasks, and report four key findings.
First, information content strongly depends on cognitive context; its absolute level and spatial distribution depend on the cognitive task.
Second, while information content shows similarities to other measures of brain activity, it is distinct from both Neurosynth maps and task contrast maps generated by a general linear model applied to the fMRI data.
Third, the brain's structural wiring constrains the cost to control its state, where the cost to transition into high information content states is larger than that to transition into low information content states.
Finally, all state transitions---especially those to high information content states---are less costly than expected from random network null models, thereby indicating the brain’s marked efficiency.
Taken together, our findings establish an explanatory link between the information contained in a brain state and the energetic cost of attaining that state, thereby laying important groundwork for our understanding of large-scale cognitive computations.

\end{abstract}
\maketitle


\section*{Introduction}
\label{sec:introduction}

\noindent
As information is processed in the brain, activity propagates between different regions along structural connections~\cite{avena2018communication,srivastava2020models}.
At the macroscopic scale in humans, these connections are white matter tracts which link one region of the brain to another~\cite{jbabdi2011tractography}; the collection of such tracts is referred to as the structural connectome~\cite{johansenberg2013human}.
The architecture of the connectome, or the pattern of tracts, determines where activity can flow~\cite{sorrentino2021structural, imms2021navigating}, and hence how the brain can move from one information-bearing state to another~\cite{ju2020dynamic}.
\LW{Depending on the task, these brain states may follow recurring patterns or show unusual activity.
The probability of the brain being in a certain state can be expressed using information content.
Yet little is known about how a state's information content depends on the brain's transition constraints.}

\begin{figure*}[ht]
    \centering
    \includegraphics[width=0.9\textwidth]{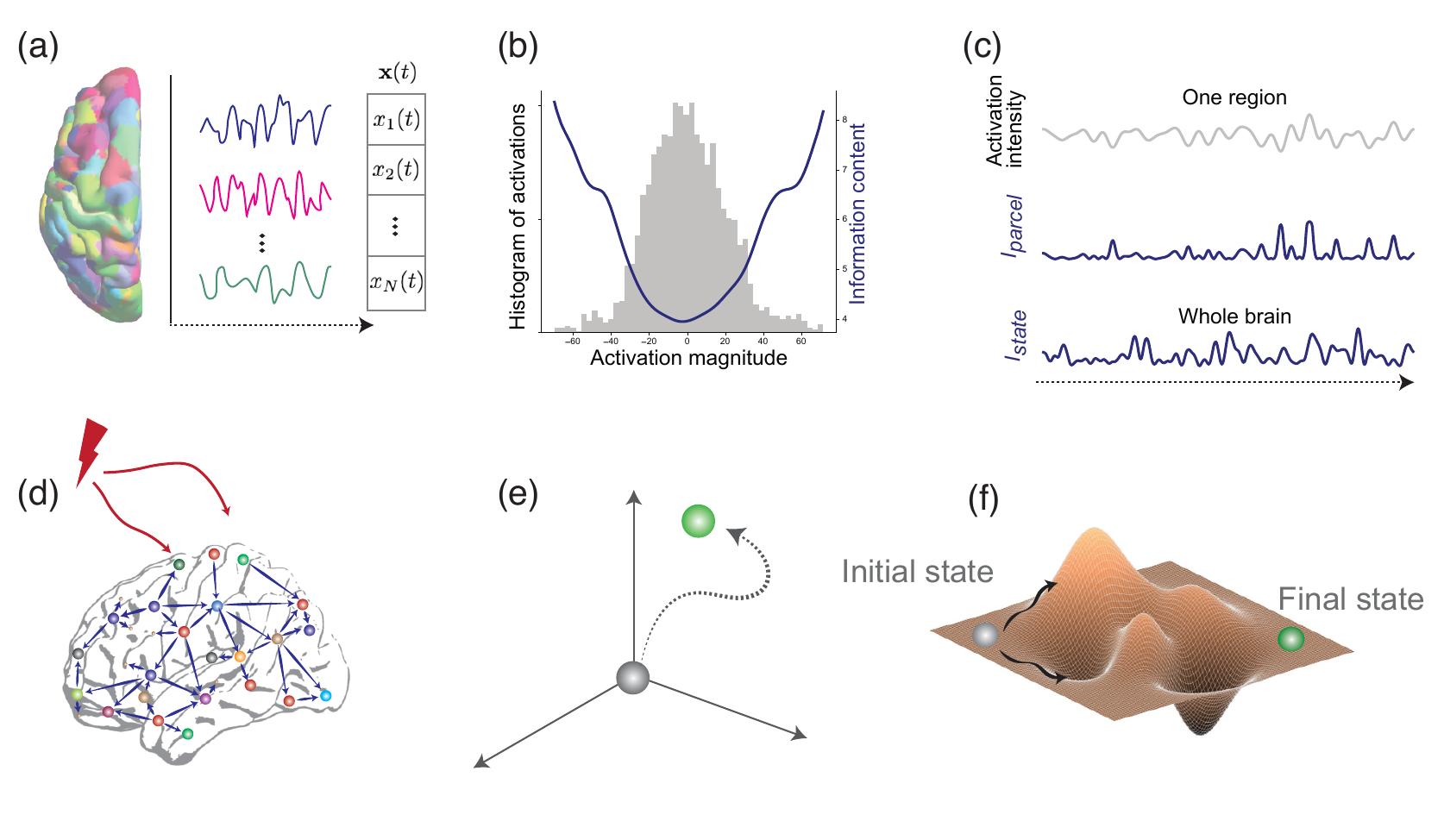}
    \caption{\textbf{Calculation of information content and transition cost of brain states.}
    (a) fMRI activation intensities of different brain \LW{parcels} define instantaneous brain state, and are represented by dots in an $N$-dimensional state space, where $N$ is the number of brain \LW{parcels}.
    (b) An example histogram of the resting state data of the \LW{parcel} showing the median information content of a single subject.
    Resting-state time series were used to estimate the probability density of activation values for each brain \LW{parcel} and each subject.
    (c) Top panel: The activation time-series of the same \LW{parcel} during the working memory task.
    Middle panel: Corresponding \LW{parcel} information content $I_{\rm{parcel}}$.
    Bottom panel: Whole brain information content $I_{\rm{state}}$, as obtained by summation over all \LW{parcels}.
    (d)-(f) By injecting control signals that diffuse on the underlying substrate of the structural connectivity matrix (schematically shown in panel (d)), the instantaneous state of the brain is moved from an initial state to a target state (e).
    (f) The cost of such a transition is determined by an underlying energy landscape and was estimated as the cost of control in the framework of network control theory.
}
    \label{fig:fig1}
\end{figure*}

Towards this understanding, prior work has made significant progress in understanding how the brain's structural connections influence the evolution of its \emph{state}, which we define as a pattern of activity across neural units, following Refs.~\cite{beynel2020structural,towlson2020synthetic,singleton2021lsd,gu2015controllability}.
Using this definition, prior work in network control theory models the macroscopic brain dynamics as a function of the structural connectome, instantiating the flow of activity along white matter tracts~\cite{gu2015controllability,towlson2020synthetic,pasqualetti2014controllability,liu2011controllability,betzel2016optimally,gu2017optimal}.
In tandem, other work has formalized conditions to define the information housed in an event based on the event's probability~\cite{shannon_info,collell2015brain}, whereby low probability events carry more information about the system than high probability events.
Hence, it is now timely to bring together these two developments and develop a novel theory of the dynamics and control of information flow across brain structure. 
Specifically, we can ask: How does the dynamics atop the structural connectome move the brain 
to high information states? What is the link between the cost of transitioning to a given brain
state and its information content?
To establish this link, we employed data from the Human Connectome Project (HCP), which includes functional magnetic resonance imaging (fMRI) and diffusion weighted imaging (DWI) scans from 596 healthy adult human participants~\cite{barch2013function}.
We represented the estimates from the DWI scans in an undirected weighted adjacency matrix encoding the structural connectome, and placed the connectome within a dynamical model to calculate the energy required to transition into low \emph{versus} high information content states (Fig.~\ref{fig:fig1}).
Collectively, these calculations provide us with the variables necessary to address our questions.

We framed our investigation by posing three primary hypotheses.
First, we hypothesized that the brain will evince different levels of information content depending on the cognitive function elicited.
This hypothesis stems from the observation that cognitive tasks constrain the brain across a pre-specified set of states, thereby changing the variability of observed activation patterns~\cite{lynn2021broken,gonzalez2021how,cohen2018behavioral}.
Second, we hypothesized that the structural connectome is organized to support transitions to observed brain states, such that the energy required for these transitions is lower than expected in random network null models.
This hypothesis stems from evidence for a bidirectional dependence between structure and function: neural units that co-activate are likely to have stronger structural connections in the future~\cite{vukovic2021rapid,sagi2012learning,ilg2008gray}, and neural units that are structurally connected are likely to more strongly co-activate in the future~\cite{wendelken2017frontoparietal}.
Hence, the brain's pattern of wiring shapes---and is shaped by---patterns of activity.
Third, we hypothesized that the energy required to reach high information content states will be greater than the energy required to reach low information content states.
This hypothesis stems from the intuition that the brain---which exists under marked energetic constraints~\cite{raichle2002appraising}---will seek to minimize the frequency with which it passes through energetically costly states.
Hence, high probability states should be those that are less energetically costly, whereas low probability states should be those that are more energetically costly.
In the process of testing and validating these hypotheses, we uncovered a generic predictor of the costs to reach observed brain states, indicating that our results are relevant to brain dynamics beyond the cognitive contexts and limited scan durations used here.
Taken together, our work integrates information theory, network control theory, and neuroimaging to demonstrate that the information content of brain states is explained by structural constraints on state energetics.

\section{Mathematical Framework}
\label{sec:math_frame}

\subsection{The information content of a brain state}
\label{sec:methods_ic}

\noindent
As formalized by Claude Shannon~\cite{shannon_info,borda2011fundamentals}, the information content of an event depends on its probability of occurrence.
Intuitively, a more surprising event carries more information than a commonplace event.
In formalizing this notion, we considered three desirable properties of a measure of information content~\cite{borda2011fundamentals}: the information content of an event must be greater than or equal to zero; absolutely certain events have zero information content; and the information content should be additive for independent events.
These properties lead to a simple definition.
Given the probability of occurrence $P_\mathbf{x}$ of a brain state $\mathbf{x}$, the information content $I_{\rm{state}}$ of that state is 

\begin{equation}
\label{eq:orig_ic}
    I_{\rm{state}}(\mathbf{x}) = \log \left( \frac{1}{P_\mathbf{x}} \right) = - \log \left( P_\mathbf{x} \right).
\end{equation}

\noindent \LW{Here, $\mathbf{x}$ is an $N$-dimensional vector, where $N$ is the number of distinct brain regions, which are referred to here as \textit{parcels}.
$P_{\mathbf{x}}$ denotes the probability with which state $\mathbf{x}$ occurs.}

To measure the information content of a given brain state, it is hence necessary to have an estimate of the probability of that state.
To determine this distribution, we first write the $N-$dimensional state vector in its components $\mathbf{x} = \{ x_i \vert i \in \{1,...,N \} \}$, where $x_i$ denotes the fMRI activation value of the $i^{\rm{th}}$ \LW{parcel} (Fig.~\ref{fig:fig1}(a)).
An estimation of the probability of a high-dimensional vector $\mathbf{x}$ would require a determination of the joint probability distribution function of all its components.
With a limited number of sample points, such a determination is numerically unstable~\cite{bishopbook}.
However, when the different entries in the random vector are independent from one another, the probability $P_\mathbf{x}$ can be written as a product of the independent probabilities of occurrence of its single entries $x_i$, making it numerically feasible to calculate the probability of a whole brain state. 
Accordingly, we disaggregated the brain into \LW{parcels} that maximize the difference of functional connectivity patterns between different \LW{parcels}~\cite{schaefer_atlas} (see Section~\ref{sec:methods}).
Making the simplifying assumption that the activities of \LW{parcels} are completely independent from one another, Eq.~\eqref{eq:orig_ic} can be reformulated as

\begin{equation}
\label{eq:ic2}
    I_{\rm{state}}(\mathbf{x}) = - \log \left( \prod_{\rm{i=1}}^{\rm{N}} p_{\rm{i}} \right) = - \sum_{\rm{i=1}}^{\rm{N}} \log \left( p_{\rm{i}} \right),
\end{equation}

\noindent
where $p_{\rm{i}}$ is the probability of the activation magnitude of a specific parcel during the observed state.
As $- \log \left( p_{\rm{i}} \right)$ can be regarded as the information content of the individual parcel $I_{\rm{parcel}}$, we write 

\begin{equation}
\label{eq:ic_parcels}
I_{\rm{state}}(\mathbf{x}) = \sum_{\rm{parcels}} I_{\rm{parcel}} .
\end{equation}

\noindent This formulation allowed us to calculate the information content associated with a single parcel $I_{\rm{parcel}}$, and thus enabled the analysis of the regional and temporal distribution of information content across the brain.
The simplifying assumption of completely independent parcels does not influence the information content of single \LW{parcels}.
\LW{However, it should be noted that the assumption leads to an overestimation of information content of whole brain states when parcels display correlated activity.
This is shown in Supplementary Material Section II, where the assumption of independence between parcels was relaxed by jointly estimating the probability distribution over pairs and quadruplets of parcels.
This approach assumes independence only between the pairs or quadruplets of parcels.
Importantly, our main results hold in these more stringent scenarios.
}

To obtain the probability distribution $p_{\rm{parcel}}$, we used the time series of fMRI activation values from a resting state scan (for an example, see Fig.~\ref{fig:fig1}(a)-(c)).
We separately estimated the probability of activation values for each subject and for each brain \LW{parcel}.
Specifically, we used a nonparametric Parzen-Rosenblatt window estimation to determine the probability density, which is a standard statistical technique for estimating the probability distribution of a random variable~\cite{mi_fmri, parzen_kde}. The approach is particularly advantageous in the estimation of a continuous unknown distribution (also see section~\ref{sec:methods}).
To ensure our method was performing as designed, we confirmed that high values of activation have low probability and are associated with high values of information content (for an example, see Fig.~\ref{fig:fig1}(b)).

The information content of activations in single \LW{parcels} during task conditions was then calculated by using the mapping of activation magnitude to information content as obtained from the rsfMRI session.
Assuming the independence of \LW{parcels}, we calculated whole brain information content by summing up over all parcels for a specific timepoint.
Further, the mean information content of a \LW{parcel} during a specific task was obtained by averaging over the information content of all timepoints.
This property can be regarded as an estimation of the cross-entropy~\cite{murphy2013machine} between the t-fMRI distribution and the rs-fMRI distribution.

\subsection{Network control theory}
\label{sec:methods_ce}

\noindent
Having calculated the information content of brain states, we now turn to the question of quantifying the structural constraints on transitions to states bearing high \emph{versus} low information content.
As noted earlier, we operationalized this question within the framework of network control theory (Fig.~\ref{fig:fig1}(d)-(f)).
In line with empirical evidence, we assumed that information moves along the structural pathways linking brain \LW{parcels}.
Following prior work~\cite{beynel2020structural,towlson2020synthetic,singleton2021lsd,gu2015controllability}, we approximated these dynamics with a linear model, which is both (i) in line with \LW{observed linearities in the fMRI blood-oxygen-level-dependent (BOLD) signal~\cite{nozari2021is} and} (ii) appropriate as a local approximation of a nonlinear dynamics occurring at finer spatiotemporal scales~\cite{colloq_cdbn}.
Coupling this intrinsic evolution with the influence of extrinsic control signals, we write down the following model (also see section~\ref{sec:methods}):  

\begin{equation}
\label{eq:lin_dyn}
    \mathbf{\dot x} = \mathbf{A}_{\rm{norm}}\mathbf{x}(t)+\mathbf{B}\mathbf{u}(t).
\end{equation}

\noindent Here, $\mathbf{x} \in \mathbb{R}^{N \times 1}$ denotes the instantaneous brain state as measured by fMRI.
The matrix $\mathbf{A}_{norm} \in \mathbb{R}^{N \times N}$ denotes the normalized structural connectome (see~\cite{karrer2019}):  

\begin{equation}
    \mathbf{A}_{\rm{norm}} = \frac{\mathbf{A}}{|\lambda (\mathbf{A})_{\rm{max}}|\times (1+c)} - \mathbf{I},  
\end{equation}

\noindent where $\mathbf{A} \in \mathbb{R}^{N \times N}$ denotes the structural connectivity matrix estimated from DWI.
The matrix $\mathbf{I} \in \mathbb{R}^{N \times N}$ denotes the identity matrix, and $\lambda (\mathbf{A})_{max}$ is the largest eigenvalue of $\mathbf{A}$.
The normalization factor $c$ was set to $0.001$ for all calculations, thereby ensuring that the dynamics in Eq.~\eqref{eq:lin_dyn} are stable.
Notably, this choice of $c=0.001$ also sets the unit of time as the inverse of $1.001 \vert \lambda(\mathbf{A}_{\rm{max}} ) \vert$.

\LW{This normalized structural connectome can be used to calculate the cost of a state transition, i.e., the effort required to drive the brain from an initial state to a final target state is calculated using control energy.}
In the $N$-dimensional state space, this external effort $\mathbf{u}(t)$ is injected into the system via the input matrix $\mathbf{B}$.
Here, we set $\mathbf{B}$ to the identity matrix, hence injecting control input into all nodes.
The energy of the state transition is then defined in terms of the external control input as $ \int_0^T \mathbf{u}(t)^\top \mathbf{u}(t) dt$.
In theory, there could exist many different control input time series that would successfully drive the brain from an initial state to a final state.
However, here we are not interested in all solutions, but instead in the solution associated with the minimum energy cost.
The reason we are interested in this minimum energy solution is that it provides us with a lower bound on the energy needed, and is consistent with our understanding of the biological constraints on energy use in the brain~\cite{raichle2002appraising,laughlin2001energy}.
To calculate the minimum control energy associated with an optimum trajectory between two states, we solved the equation

\begin{equation}
    E_{\rm{min}} = \underset{\mathbf{u}}{\text{min}}  \int_0^T \mathbf{u}(t)^\top \mathbf{u}(t) dt
\label{eq:e_min}
\end{equation}

\noindent constrained by Eq.~\eqref{eq:lin_dyn}, the initial state $\mathbf{x}(0)$ and the final state $\mathbf{x}(T)$.
\LW{Details on the solution for the present case are given in Appendix~\ref{sec:app_ed_mce}; for a more general solution the reader is referred to Ref.~\cite{colloq_cdbn}.}
We calculated $E_{\rm{min}}$ for a given brain state by setting it as the final state at time $t=T$ while the initial state was set to the origin (i.e., the mean brain state in the data).
We select $T=10$ in order to choose a time horizon regime with higher order network effects that still shows high correlation to other time horizons~\cite{karrer2019}.

Our estimates of the energy required for specific transitions were constrained to the set of states that we can observe in the fMRI data.
However, it is interesting to ask whether there exists some generic characteristic of the brain that both (i) explains the observed states and (ii) enables us to make predictions about the set of states that we may not have observed in the short imaging sessions that are well-tolerated by human participants.
One reasonable candidate for this generic characteristic is the brain's average controllability, which quantifies the ease of moving the instantaneous state of the brain to any location on the \PS{underlying energy landscape that governs the cost of transition between brain states~\cite{karrer2019}. This, in turn, provides an estimate of the accessible volume in the state space.}
Indeed the quantity of average controllability is the measure of average energy required to reach random states uniformly distributed on the surface of a hypersphere of unit radius~\cite{avg_controllability1}.
If a brain has low average controllability, intuitively it should require more energy to reach observed states as present in the fMRI data, but also to reach unobserved states that were not measured.
The average controllability was calculated for state equations with stable dynamics as the trace of the controllability Gramian $W_C$, defined as: 
\begin{equation}
    \mathbf{W_C} = \int_0^T \mathrm{e}^{\mathbf{A}t} \mathbf{B} \mathbf{B}^\top \mathrm{e}^{\mathbf{A}^\top t} dt.
\end{equation}
\noindent In our investigation, we used this calculation to test our intuition that the brain's average controllability explains the energy required to reach \emph{observed} states, and to predict the energy required to reach \emph{unobserved} states, thereby augmenting the potential relevance of our study to other contexts.

\section{Results}
\label{sec:results}

\noindent
Striking a balance between the requirement of sufficient information and the limitations of fMRI acquisitions, we set the number of parcels $N$ to be 100 in our experimental protocol.
All experiments were repeated with a resolution of 400 parcels (see Supplementary Material Section II).
Having estimated the probability distribution of brain states using the resting state fMRI data (see Section~\ref{sec:math_frame}), we first calculated and compared $I_{\rm{state}}$ across different tasks.

\subsection{Levels of information content for different cognitively demanding states}

\begin{figure}[ht]
    \centering
    \includegraphics[width=.45\textwidth]{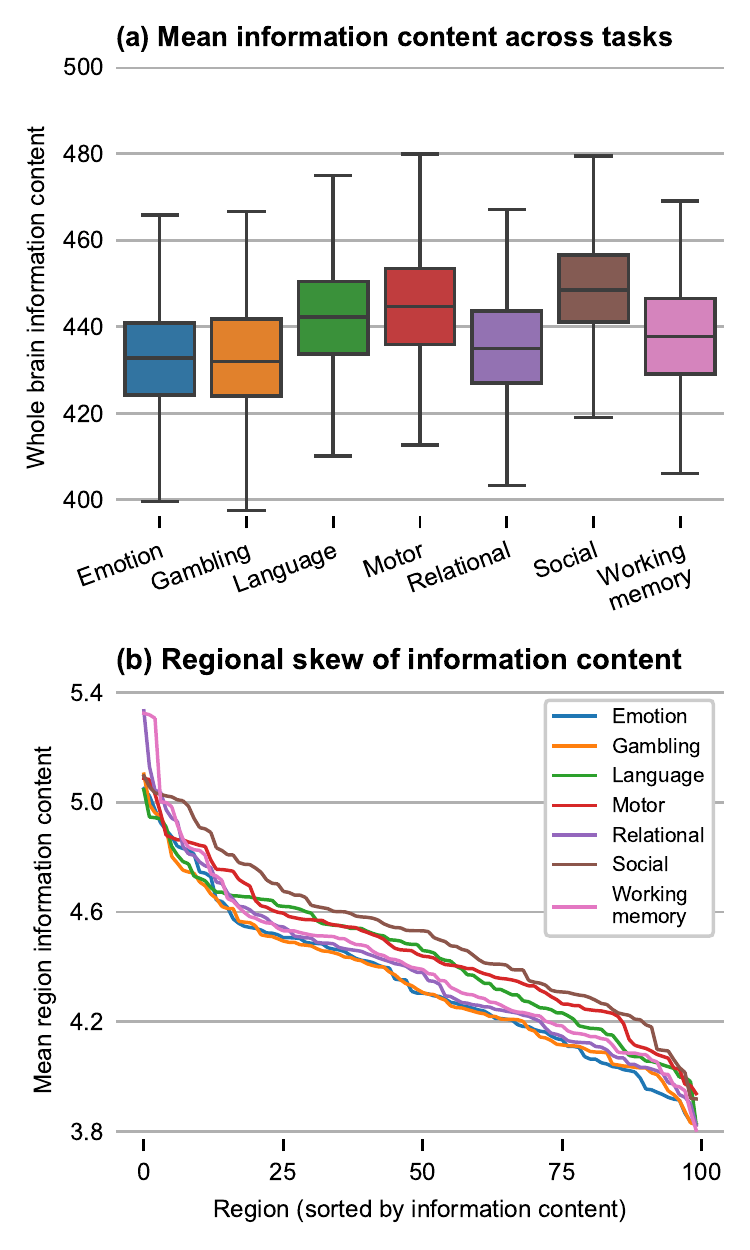}
    \caption{ \textbf{Information content across different tasks.}
    (a) Mean information content. Among all seven tasks considered in this work, the brain exhibited highest overall information content during the social task (p\textless0.001 after Bonferroni correction for multiple comparisons), and the lowest information content in the emotion and gambling tasks.
    (b) Regional skew of information content. For the different tasks, the mean \LW{parcel} information content $I_{parcel}$ across all subjects was calculated.
    The \LW{parcels} were then sorted by $I_{parcel}$ for every task, and the resulting declining lines were plotted for every task.
    }
    \label{fig:fig2_mean_info_content}
\end{figure}

\noindent
To determine the task-dependence of information content, we analyzed the mean information content for all observed task states in all 596 subjects.
The different task settings, emotion, gambling, language, motor, relational, social, and working memory in the HCP dataset~\cite{hcp_tfmri} exhibited distinctively different levels of information content (see Fig.~\ref{fig:fig2_mean_info_content}(a)).
The social task showed significantly higher information content than all other tasks (p\textless0.001).
In contrast, the mean $I_{\rm{state}}$ of the gambling and the emotional task were lower than for any of the other tasks.

After an estimation of the mean information content of the global brain state, we next investigated the distribution of information content across brain \LW{parcels}.
The distribution of $I_{\rm{parcel}}$ across brain \LW{parcels} was not significantly different from a Gaussian distribution (verified by a Shapiro-Wilk test, p\textgreater.05) for the gambling, language, motor and social tasks.
However, the Gaussian assumption did not hold for the other three tasks, namely working memory, emotion, and the relational task, which showed positive skew (p\textless0.05).
This finding indicates that the regional distribution of information content in these tasks was anisotropic, with some \LW{parcels} contributing disproportionately more than others.
As shown in Fig.~\ref{fig:fig3_approaches}(b), information content was heavily concentrated in only a limited part of the brain during those three tasks.
In the next section, we map the regional distribution of information content and compare it to other measures derived from fMRI activity.

\subsection{Comparing information content to other fMRI assessments}

\begin{figure*}[ht]
    \centering
    \includegraphics[width=0.9\textwidth]{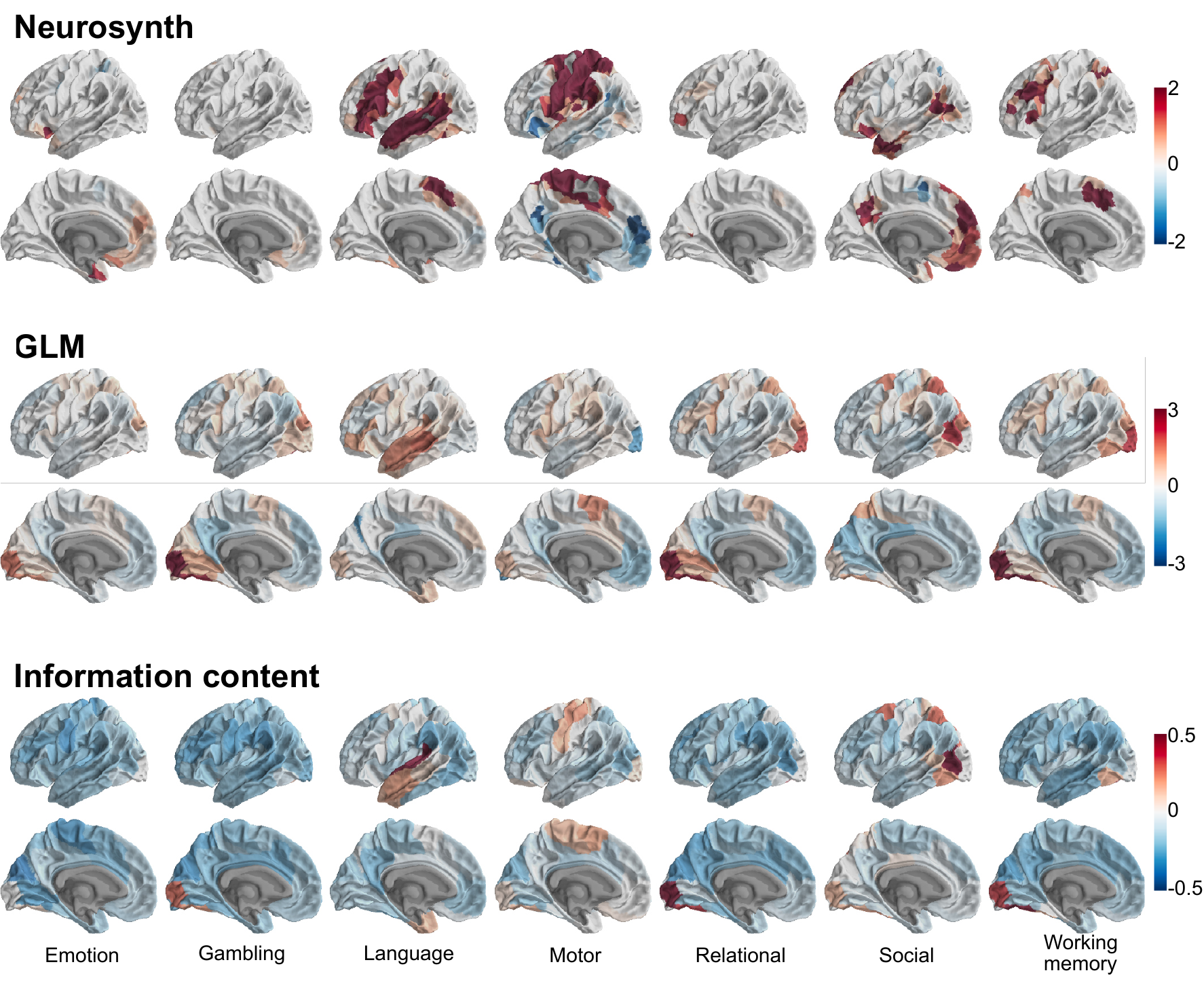}
    \caption{\textbf{Regional distribution of three different measures derived from fMRI activation.} Brain parcels with involvement for the different tasks according to community consensus (Neurosynth~\cite{neurosynth}, top), a classical generalized linear model approach (GLM, middle), as well as the proposed information content metric (bottom).
    \LW{For information content, the difference between information content of the task fMRI distribution and the rsfMRI base distribution, averaged over all subjects and time points, is visualized.}
    Red: positive values, blue: negative values.
    All non-significant \LW{parcels} (p\textgreater0.01 after Bonferroni correction) were set to zero and colored in gray.
    As the results are approximately symmetric across the hemispheres, only the left hemisphere is visualized here.
The right hemisphere can be found in Supplementary Materials Section III.}
    \label{fig:fig3_approaches}
\end{figure*}

\noindent
In this section, we sought to understand the task-dependence of the regional distribution of information content, and to compare it to previously defined measures of activity in fMRI data.
First, we compared regional profiles of information content with the consensus in the field on which brain regions are involved during specific tasks using data from a large-scale meta-analysis.
For this purpose, the keywords of the different tasks were examined in Neurosynth~\cite{neurosynth}, which searches published neuroimaging articles for specific keywords.
From these articles, brain coordinates were automatically synthesized.
Second, we used the classical generalized linear model (GLM) method to identify activated \LW{parcels} in the same dataset as used for analysis of information content.
\LW{Specifically, a linear model without intercept was used to predict task-based activation for each subject and each parcel.
In this model, the predictor is derived from the task blocks.
During fixation and initiation blocks, the predictor is set to zero.
It is set to one during all task blocks.
This predictor is then regressed against the actual fMRI activation.
Finally, the 100 coefficients (one for each parcel) are \textit{z}-normalized per subject, and the \textit{z}-scores are averaged over all subjects to obtain the final GLM score.}

In Fig.~\ref{fig:fig3_approaches}, these two approaches are displayed next to the information content metric.
While for some tasks, the three metrics showed similar profiles across brain \LW{parcels}, other tasks exhibited strongly diverging effects.
Clear correspondence was observed during the language task: the middle and superior temporal gyri are distinctively marked in all three approaches.
We note the difference in the inferior frontal gyrus between the Neurosynth results and the remaining two methods that used the HCP data.
This strong difference could have arisen from the fact that the HCP language task consisted only of language comprehension and excluded language production, which is not reflected when using the keyword ``language'' in Neurosynth.

\begin{figure*}[hbt]
    \centering
    \includegraphics[width=1\textwidth]{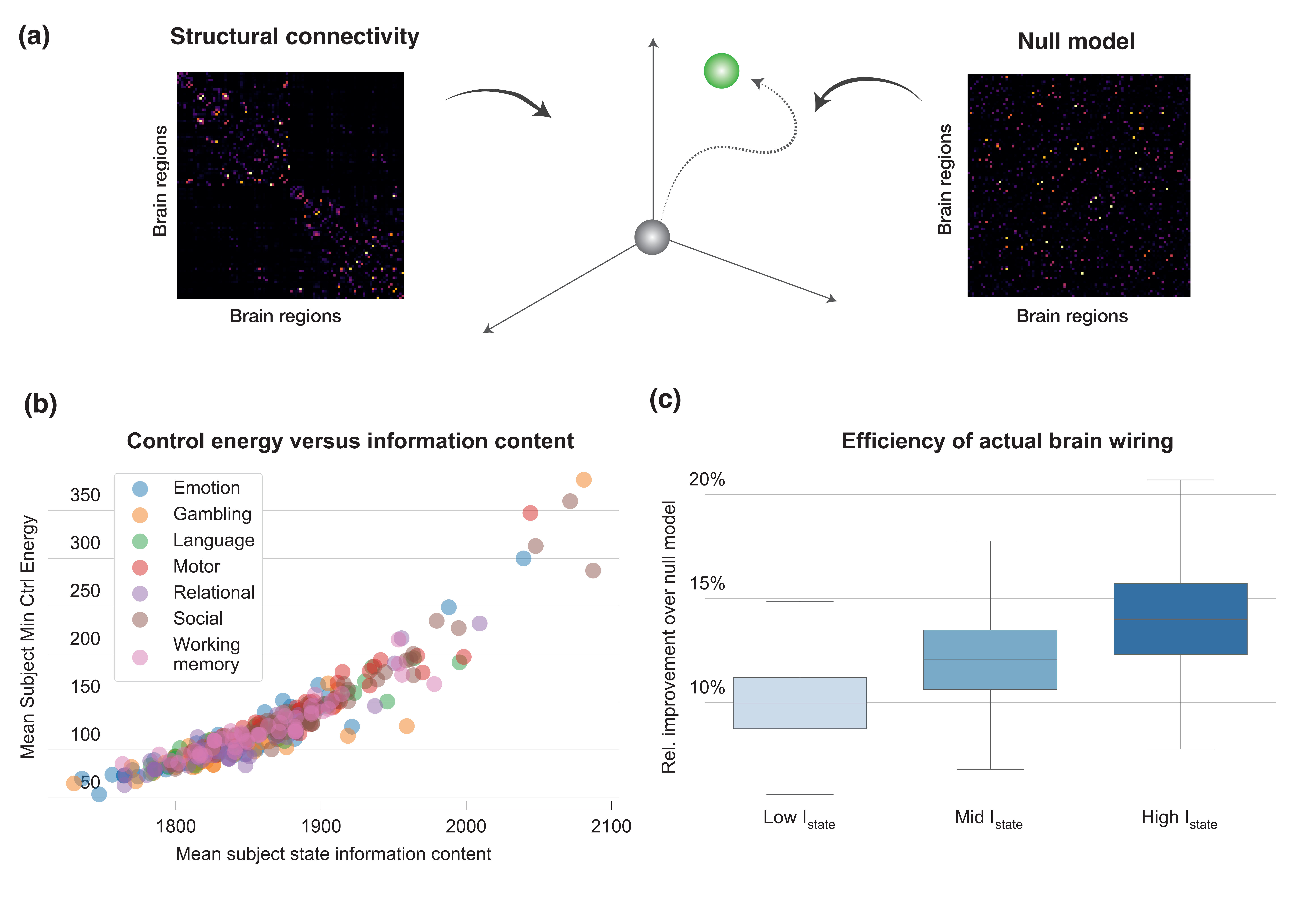}
    \caption{\textbf{Energetics of information content.}
    (a) The scheme of calculating the improvement in the cost of transition using the actual structural connectivity (\PS{on the left}) of the brain over that calculated using a null model (\PS{on the right}). \PS{Heatmaps show an example structural connectivity matrix (on the left) and a randomized null model (on the right)}. 
    The minimum control energy to reach all the observed brain states calculated using each model was then compared with the other to calculate the improvement. 
    (b) Mean state information against mean minimum control energy for all subjects and all tasks in the HCP dataset.
    For an uncluttered representation, only every 50th datapoint is plotted.
    (c) Efficiency of actual brain wiring. Comparison of real-world control energy versus control energy of connectivity null models for high, middling, and low information content states.
    }
    \label{fig:fig4_energetics}
\end{figure*}

The activation profile for the motor task also showed strong overlap for all three approaches, highlighting the human motor cortex.
Compared to the information content approach, where the motor cortex was the only marked area, the GLM noted activation in the visual cortex, while Neurosynth also pinpoints previous findings in the inferior temporal gyrus.
For other tasks, the three metrics showed more diverging profiles.
Both metrics on the HCP dataset often highlight the visual cortex, e.g., during gambling, relational processing, or the working memory task.
In general, the regional profiles of GLM and information content showed more similarities to each other than to the profile obtained from Neurosynth.
However, the involvement of certain frontal lobe areas, as found in Neurosynth as well as with the GLM, cannot be seen with information content.
In contrast, the information content approach generally highlights fewer areas than the GLM, indicating a higher specificity. Broadly, these findings serve to highlight the fact that information content is a complementary and not redundant metric. 

\subsection{Brain architecture minimizes energy requirements for high information states}

\noindent
In the previous section, we found that the brain exhibits different levels and patterns of regional information content depending on the task condition.
We then asked how the cost of transition to the task states relates to their information content.
Accordingly, we examined the fMRI states observed in all tasks, and calculated $E_{\rm{min}}$ as defined in Eq.~\eqref{eq:e_min}) for all states.
The initial state was set to the mean state, which coincides with a state of zero activation.
Since the probability distribution is unimodal (see Fig.~\ref{fig:fig1}(b)), high information content states are naturally far away from the mean state and thus should require higher energies than states with low information content (see Appendix \ref{sec:app_ed_mce}).
While this correlation was noisy for single states, it was clearly visible when analyzed on a per-subject-level;
subjects with higher $I_{\rm{state}}$ required much higher $E_{\rm{min}}$ (Fig.~\ref{fig:fig4_energetics}(a)).
This finding taken together with the observation that activations with high information content appear in the tails of the unimodal probability distribution and are observed with lower frequency (see Fig.~\ref{fig:fig1}(b)), implies that a higher energy is required to drive the brain into states that are observed less frequently.
This is a key point that provides support for the framework of network control theory, and its particular instantiation used here. Subjects that reach rare brain states during a given task---and thus exhibit high levels of information content---also exhibit higher control energy.
Thus, the quantity $E_{\rm{min}}$ captures the inherent difficulty of state transitions atop the underlying structural connectivity during the performance of cognitively demanding tasks.

Having established the above connection between $E_{\rm{min}}$ and $I_{\rm{state}}$ of subjects, we further sought to examine whether the structural connectivity of the brain network is inherently suited to make transitions to states with high information content.
Accordingly, we calculated the average improvement in control energy calculated using the measured structural connectivity matrix over that calculated using a randomized model of brain connectivity that preserved the original degree, weight, and strength distributions (see Section~\ref{sec:methods_ce}).
We calculated these improvements for states with low, intermediate, and high $I_{\rm{state}}$ for each subject.
\LW{The states with low, intermediate, and high $I_{\rm{state}}$ were retrieved by sorting the brain states from all tasks together based on their $I_{\rm{state}}$, and were separated into three equal parts for each subject.}

The improvement was then defined as the mean difference in control energy for the two models for the three sets of states.
We found that the null model required higher control energy for all three types of states.
However, the improvement in transitioning cost when using the actual structural connectivity over that using the null model was significantly higher for high information content states in comparison to low information content states (Fig.~\ref{fig:fig4_energetics}(b), one-way ANOVA test with p\textless 0.001).
Thus, while the actual architecture of the structural connectome performed better for all three ranges of information content, it was particularly suited to reach high information content states when compared to a null model.

\subsection{Covariates of minimum control energy and information content}

\noindent
In a final set of analyses, we examined possible covariates between information content and minimum control energy, and their influence on these two approaches.
A major covariate that influences both minimum control energy and information content is the Euclidean distance between the observed state and the mean state.
Naturally, the further away from the mean state, the higher the $E_{\rm{min}}$.
In the absence of any network effects, the minimum control energy is equal to the squared Euclidean distance of the brain state (for a theoretical derivation, see Appendix \ref{sec:app_ed_mce}).
Similarly, if fMRI activations were Gaussian distributed with no difference within and between subjects, $I_{\rm{state}}$ would be linearly related to the squared Euclidean distance (Appendix \ref{sec:app_ed_ic}).
In reality, fMRI activations vary both within and between subjects~\cite{fMRI_amp_fluct}, and network effects are non-negligible in control systems.
To evaluate whether $I_{\rm{state}}$ predicts the minimum control energy $E_{\rm{min}}$ beyond the squared Euclidean distance in the real data, we tested a linear model with only squared Euclidean distance as input against a linear model containing both squared Euclidean distance and $I_{\rm{state}}$ as input.
The model containing both features showed better predictiveness, with a mean absolute difference of 13.2, against 24.0 for only squared Euclidean distance \LW{(p\textless0.001, Wilcoxon signed-rank test)}.
\LW{Scatter plots between $I_{\rm{state}}$ and $E_{\rm{min}}$, as well as between squared Euclidean distance and $E_{\rm{min}}$ are shown in Supplementary Material Section III.}

Another covariate that could explain the individual differences in minimum control energy is average controllability.
Average controllability offers insights into the ease of navigating the state of brain \LW{parcels} along all the directions in the state space on average by external control signals~\cite{gu2015controllability}.
It is thus independent of specific brain states, and with the control set fixed across all subjects, it is determined solely by the underlying structural connectivity.
We evaluated whether average controllability could explain the difference between $E_{\rm{min}}$ as predicted from $I_{\rm{state}}$, and actually measured $E_{\rm{min}}$.
We hypothesized that if a brain system had high average controllability, the minimum control energy should be overestimated.
Similarly, if a brain showed low average controllability, the minimum control energy, as predicted from $I_{\rm{state}}$, should be underestimated.
Thus, the error between the predicted $E_{\rm{min}}$ and the computed $E_{\rm{min}}$ should show positive correlation with the average controllability: for high average controllability an overestimation, while for low average controllability an underestimation.
Verifying this hypothesis using the Pearson correlation coefficient, we found that the error between the predicted and computed $E_{\rm{min}}$ is positively correlated with average controllability (Pearson correlation coefficient of 0.09, p\textless 0.05).
In other words, for subjects with high average controllability, $E_{\rm{min}}$ is overestimated if it is predicted from $I_{\rm{state}}$---subjects with high average controllability require less effort to reach high information content states than subjects with low average controllability.
\LW{A visualization of this correlation and details on the calculation procedure can be found in Supplementary Material Section III.}
Thus, the theoretical measure of average controllability indeed translates into the ease of reaching real-world brain states.

\section{Discussion}
\label{sec:discussion}

\noindent
We defined the measure of information content associated with each activation value, inferred from the resting state probability distribution.
This measure allowed us to determine the information content of each brain \LW{parcel} as well as that of whole brain states, and led us to compare information content across different tasks. 
For example, the highest whole brain information content was found in the social task.
Meanwhile, the middle and superior temporal gyri showed high $I_{\rm{parcel}}$ during the language task, and the motor cortex during the motor task.
Collectively, information content is sensitive to task differences across spatially distributed brain \LW{parcels} and could be used to evaluate the relationship between altered brain network dynamics, self-referential processing, and executive dysfunction in psychopathology~\cite{whitfield2012default}.

Brain state probabilities provide a useful measure of information content that is consistent yet non-redundant with prior fMRI assessments and is strongly correlated with theoretical linear dynamical models of functional activity trajectories.
This empirical correlation distinguishes the present approach from prior work applying information theory metrics to brain function, structure, and activity variance~\cite{ito2020task, amico2020information}, providing a conceptual link between information theory and control theory for efficient network communication~\cite{ponce2015task,srivastava2020models}.
Other information theoretical measures in neuroscience have been mostly applied to obtain insights into nonlinear relationships of multivariate data or to quantify uncertainty~\cite{itn_tutorial, it_reinagel, rabinovich2012principles}.
For example, an information theoretical concept of cognitive control has been proposed~\cite{itacc_fan}, a robust information flow metric for the human brain is available~\cite{est_flow_info}, and the communication dynamics in brain networks have been studied~\cite{amico2020information}.
The relationship between brain entropy and human intelligence remains a particularly active area of research~\cite{e22090917, entr_intel18, bentr_fmri14, icep_liu}.
Most closely related to our approach is the application of mutual information to task fMRI data, a technique used to establish relationships between an experiment and individual voxels~\cite{mi_fmri} or to reveal the functional connectivity between \LW{parcels}~\cite{mi_fmri_knn}.
However, with previously used metrics such as entropy or mutual information, an analysis of brain states or a time-dependent analysis of information was not possible.
In contrast, information content---as used in this work---allows insights into the current condition of single brain \LW{parcels} as well as into whole brain states, and makes it possible to relate brain states to their associated transition costs.

If the control energy is a metric characterizing cognitive effort and control efficiency~\cite{gu2015controllability, tang2017developmental, braun2019brain}, and the brain develops under evolutionary constraints of efficiency~\cite{bullmore2012economy}, then states that require greater control energy to enact should be less likely to occur empirically.
Consistent with this hypothesis, we found that states that require greater control energy to enact are less likely to occur and tend to convey more information.
More probable activity states require reduced control energy.
The greater control energy cannot be simply explained by the magnitude of the fMRI activation, even if $I_{\rm{state}}$ is related to the squared Euclidean distance under oversimplified assumptions (see Appendix~\ref{sec:app_ed_mce}).

\LW{Information content has two different applications: optimal message passing and an expression of probability. 
Regarding probability in terms of information content has mathematical advantages. 
These advantages were exploited in the present paper. 
For example, “high information” was used to quantify unusual brain activity. 
However, the findings are also relevant for message passing, as they are consistent with the efficient coding hypothesis.
}

A general principle of efficient coding is to assign the shortest and least costly codes to the most frequent symbols~\cite{mackay2003information}.
This principle resembles Zipf's law, in which the most frequent states require the least effort, and which has been observed across biological and human made systems~\cite{zipf2016human}.
In our results, the most frequently used symbols are analogous to the low information content (high probability) neural states, while the most costly codes are analogous to the greater control energy cost.
Analogous to the Boltzmann distribution, the brain network will be in certain states as a function of that state's energy; states requiring lower control energy have a higher probability of being occupied.
Hence, the macroscale brain network may be subject to similar constraints of efficient coding as the microscale~\cite{bullmore2012economy, zhou2020efficient}.

Moreover, our findings are consistent with widespread observations that the human brain is efficiently wired.
We showed that, especially for high $I_{\rm{state}}$, the brain necessitates significantly less control energy than a randomized null model.
Network control theory posits a dynamical system atop the structural connectome, in which all brain \LW{parcels} are involved in the process of enacting functional activity state transitions to support future behavior~\cite{gu2015controllability}.
The observation that functional activity states requiring high control energy were less probable suggests a role of the structural connectome in constraining high-cost states.
Indeed, evolutionary processes select for brain network biology and topology under constraints of efficiency and adaptability~\cite{bullmore2012economy, ercsey2013predictive, stiso2018spatial}.
\LW{Efficient and adaptable brain dynamics across species could be characterized by controllability~\cite{kim2018role}, and recent evidence demonstrates clear empirical links between control energy and biological metabolism~\cite{control_metabolism}.
As control energy, in turn, was shown to be linked to information content, it can be hypothesized that the brain's metabolism is a driver of the discrepancy between the results of information content and GLM shown in Fig.~\ref{fig:fig3_approaches}.}

Our findings motivate future research on the coupling between structural and functional brain networks using process models of dynamic brain activity~\cite{avena2018communication}.
Recent work suggests that models of brain network communication can both improve the descriptive statistical coupling between structure and function as well as provide insight into the theoretical processes underlying brain network dynamics~\cite{suarez2020linking}.
We reported that $I_{\rm{state}}$ predicts the control energy required to enact the state.
The remaining unexplained variance of control energy, a metric derived from a dynamical network model incorporating both structure and function, was partly explained by average controllability, a metric solely based on structural connectivity.

Hence information content, minimum control energy, and average controllability represent a characterization of structure and function which provides new avenues to study and interpret brain networks as dynamical systems.
These links between information theory and the network control framework align with reports of trial-by-trial variability of neural activity decreasing during task performance to tighten task-relevant dynamical trajectories and overcome endogenous noise activity unrelated to the task~\cite{he2013spontaneous, ponce2015task, mazzucato2016stimuli, ito2020task}.
Network control theory further integrates these conceptual and methodological notions in a simpler linear dynamical model, providing both conceptual insights into and methodological quantification of the brain network.
 
Several limitations of using information content for analysis of fMRI data should be noted.
Most importantly, for analysis of task-fMRI data, an extensive rsfMRI acquisition of the same subject in the same scanner is necessary in order to estimate the probability density function of the individual brain.
Thus, the present approach is mostly suitable to studies that include an extensive rsfMRI session next to a battery of task fMRI sessions.
Further, as the calculation of $I_{\rm{parcel}}$ is based on the probability estimation for single parcels, and $I_{\rm{state}}$ is obtained by summing $I_{\rm{parcel}}$ over all parcels, $I_{\rm{state}}$ relies on the assumption of independence of parcels.
Although the parcels were obtained by a functional parcellation, the different time series are not completely independent.
Relying on this assumption could have led to a slight overestimation of information content in the human brain when different parcels are synchronized or correlated.
Further, using a linear control model represents a simplification of the neural system.
The same applies to parcellations of the human brain, and modeling brain connectivity with diffusion MRI tractography.
\LW{Lastly, state transitions were defined as going from the mean state to task states. 
This setup limits the applicability to scenarios where transitions between task states are measured, e.g., following subsequent states during a task-fMRI session.}

\section{Conclusion}
\label{sec:conclusion}

\noindent
Based on Shannon's fundamental work on entropy, we investigated information content in the human brain across subjects and mental tasks.
Different tasks demonstrate different absolute levels as well as distinct regional distributions of information content.
Compared to other tasks, the highest information was obtained during social cognition, while emotion and gambling showed relatively low levels of total information content.
In a next step, we then linked information content to control energy, showing that high information states necessitate high levels of effort to reach.
However, the brain seems to be efficiently wired:
When compared to a brain connectivity null model, significantly less energy is required with the actual brain wiring, especially for high information states. Our work generally provides an explanatory link between information content, state transition costs, and neural processing supporting cognitive function.

\section{Methods and Evaluations}
\label{sec:methods}

\subsection{Imaging data acquisition and preprocessing}
\label{sec:methods_preproc}

\noindent
The data includes functional magnetic resonance imaging (fMRI) and diffusion weighted imaging (DWI) scans from 596 healthy adult human participants (Fig.~\ref{fig:fig1})~\cite{barch2013function}.
From the fMRI data, we estimated regional and whole brain information content in seven different executive conditions: an emotion processing task, a language task, a motor task, a relational task, a social task, and a working memory task.
From the DWI scans, we extracted estimates of the strength of white matter tracts connecting the brain \LW{parcels}: primarily 100 large scale areas (see Main), and in a complementary investigation 400 smaller scale areas (see Supplement).
All analyses were performed in accordance with the relevant ethical regulations of the WU-Minn HCP Consortium Open Access Data Use Terms.

The acquisition parameters for each data type were as follows.
The parameters for the acquisition of the high-resolution structural scan were TR\,=\,2400\,ms, TE\,=\,2.14\,ms, TI\,=\,1000\,ms, flip angle\,=\,8°, FOV\,=\,224\,×\,224\,mm, voxel size\,=\,0.7-mm isotropic, BW\,=\,210 Hz/Px, and acquisition time\,=\,7:40 min.
Functional magnetic resonance images were collected during both rest and task with the following parameters: TR\,=\,720\,ms, TE\,=\,33.1\,ms, flip angle\,=\,52°, FOV\,=\,208\,×\,180\,mm, matrix\,=\,104\,×\,90, slice thickness\,=\,2.0\,mm, number of slices\,=\,72 (2.0-mm isotropic), multifactor band\,=\,8, and echo spacing\,=\,0.58\,ms.
Diffusion tensor images were collected with the following parameters: TR\,=\,5520\,ms, TE\,=\,89.5\,ms, flip angle\,=\,78°, refocusing flip angle\,=\,160°, FOV\,=\,210\,×\,180, matrix\,=\,168\,×\,144, slice thickness\,=\,1.25\,mm, number of slices\,=\,111 (1.25-mm isotropic), multiband factor\,=\,3, echo spacing\,=\,0.78\,ms, and b values\,=\,1000, 2000, and 3000\,s/mm2.

From the fMRI data, resting-state, working memory, language, relational, social cognition, emotion, gambling, and motor scans were analyzed.
We chose to use all scans in order to sample the brain’s functional activity in a diverse range of cognitive states.
Details of the task designs can be found in Ref.~\cite{hcp_tfmri}.
Brains were normalized to fslr32k via the MSM-AII registration.
CompCor, with five principal components from the ventricles and white matter masks, was used to regress out nuisance signals from the time series.
In addition, the 12 detrended motion estimates provided by the Human Connectome Project were regressed out from the regional time series.
The mean global signal was removed and then time series were bandpass filtered from 0.009 to 0.08 Hz~\cite{baker52, smith53}.
Finally, frames with greater than 0.2\,mm frame-wise displacement or a derivative root mean square (DVARS) above 75 were removed as outliers.
Segments of less than five uncensored time points were also removed.
Sessions composed of greater than 50 percent outlier frames were not further analyzed.
The processing pipeline used here has previously been suggested to be ideal for removing false relations between connectivity and behavior~\cite{opt_mod_mcnr}.
\LW{For all subjects, we parcellated the brain into 100 as well as into 400 parcels.
As our approach to whole brain information content assumes independence of parcels, the parcellation must be chosen to minimize mutual information between the parcels.
This corresponds to a minimization of correlation, i.e., of functional connectivity between the parcels.
Thus, the Schaefer parcellation scheme~\cite{schaefer_atlas} was chosen as it clusters voxels based on local-global functional connectivity.}
The two fMRI acquisitions with opposing MRI phase encoding gradient were concatenated for each fMRI timeseries.
To further reduce the effects of outliers to the probability density estimation, fMRI activations were clipped to four times their standard deviation.

For preprocessing of the diffusion imaging data, the Human Connectome Project applied intensity normalization across runs, the TOPUP algorithm for EPI distortion correction, the EDDY algorithm for eddy current and motion correction, gradient nonlinearity correction, calculation of gradient b-value/b-vector deviation, and registration of mean b0 to native volume T1w with FLIRT.
BBR+bbregister and transformation of diffusion data, gradient deviation, and gradient directions to 1.25-mm structural space were also applied.
The brain mask was based on the FreeSurfer segmentation.
The BedpostX (Bayesian Estimation of Diffusion Parameters Obtained using Sampling Techniques) output was then calculated, where the ``X'' stands for modeling crossing fibers.
Markov Chain Monte Carlo sampling was used to build probability distributions on diffusion parameters at each voxel.
The process creates all of the files necessary for running probabilistic tractography.
Using the Freesurfer recon-all data computed by the Human Connectome Project, the fsaverage5 space cortical parcellation was registered to the subject’s native cortical white matter surface, and then transformed to the subject’s native diffusion volume space.
From these data, we derived seeds and targets for probabilistic tractography, which we ran with the FSL probtrackx2 algorithm using 1000 streams initiated from each voxel in a given parcel.
This DWI pipeline was previously used in Ref.~\cite{murphy_2020}.

After performing probabilistic tractography, we applied the same Schaefer atlas as applied to the fMRI data.
Next, we calculated the proportion of streams seeded in a voxel in one parcel that reached another \LW{parcel}.
We chose to use the proportion of streamlines to represent structural connectivity due to the inhomogeneity of the \LW{parcel} sizes.
We collated all interregional estimates of structural connectivity into a single 100\,$\times$\,100 (respectively, 400\,$\times$\,400) connectivity matrix, which we then treated as the formal encoding of a network representation of brain structure~\cite{numnn}.
Similar to the model of brain function, in this structural network representation, \LW{parcels} are represented by network nodes, and structural connections are represented by weighted edges, where the weight of the edge between node $i$ and node $j$ is given by the proportion of streams seeded at parcel $i$ that reach parcel $j$.
\LW{
    Due to the imprecise natur of the diffusion tractography algorithm, slight asymmetries emerge as artifacts in the final raw connectivity matrices.
    We thus follow the standard procedure of setting the diagonal to zero, and symmetrize the connectivity matrices by setting the weight from $i$ to $j$ to be equal to the weight from $j$ to $i$.
}
As only data from subjects where all fMRI as well as DWI scans remained after scrubbing, data from 596 subjects was used in this study.

\subsection{Probability density estimation}

\noindent
For calculating the information content of the fMRI activation, a probability density estimation on the first rs-fMRI dataset was necessary.
For this purpose, a nonparametric Parzen-Rosenblatt window estimation was used.
\LW{The Parzen-Rosenblatt estimator centers a kernel function $K$ on each observed data point, sums the different kernel functions and normalizes the joint density function by the number of observed data points $N$~\cite{bishopbook}:
\begin{equation}
    p(\mathbf{x}) = \frac{1}{N}\sum_{n=1}^N K(\frac{\mathbf{x}-\mathbf{x}_n}{h}) ,
\end{equation}
where $\mathbf{x}_n$ are the observed data points, and $h$ the bandwidth, a critical parameter to set when using this technique.
As fMRI activations are unimodal~\cite{Fox05, Damoiseaux06} and a Gaussian kernel was used, it was possible to employ a common rule-of-thumb for the optimal bandwidth~\cite{kde_bandwidth}.
Using this rule, the bandwidth $h$ was set individually for each parcel and each subject as follows:
\begin{equation}
    h = \left( \frac{4 \sigma^5}{3m}\right)^{\frac{1}{5}} ,
\end{equation}
where $\sigma$ is the standard deviation of the time series, and $m$ is the number of time points in the series.
Whereas $m$ was 2400 for all subjects, $\sigma$ was independently set for every subject.}
Using the estimated probability distribution, it was straightforward to map activation intensities to probabilities and to finally calculate the information content.

\subsection{Network null model}

\noindent
To compare the measured brain connectivity against a randomized version of possible neural connections, a network null model was utilized.
It is useful for such as null model to retain certain features of the real-world graphs to ensure that the comparison is precise and isolates biological factors of interest.
Here, network null models that have the same degree, weight, and strength distributions as the original graph were constructed.
The following rules were applied to obtain these random matrices:
With a chance of 50\%, all columns were randomly permuted, and the permutation order was retained.
Using the exact same permutation, all rows were then permuted.
In the other 50\% of cases, the rows were first randomly permuted, and then the columns switched in the same order.
This procedure was repeated 1000 times.
This random permutation naturally does not change the degree, weight, or strength distributions, as either only complete rows switch place, or the entries in a row are shuffled in the row.
By using the same permutation for rows and columns, the zero diagonal could be retained.

\subsection{Comparison to other approaches}

\noindent
To compare the regional results of information content during task to other methods of task-based fMRI analysis, two different approaches were chosen.
The first approach consisted of using Neurosynth~\cite{neurosynth} to obtain a meta-analysis of previous studies on the same topics, and to identify the locations reported in those studies.
The second approach consisted of using a generalized linear model (GLM), the most commonly utilized task-based fMRI approach~\cite{hitchhiker_fmri}, and also the tool used in the original analysis of the HCP task fMRI data~\cite{hcp_tfmri}.
For Neurosynth, the terms working memory, language, relational, social cognition, emotion, gambling, and motor were searched using the provided toolbox.
From the results, the association test maps were extracted with the default settings (expected false discovery rate of 0.01). To obtain a GLM model of the HCP data with comparable input as the information content analysis, the design matrix was created by setting all task blocks to one, and all fixation and initiation blocks to zero.
The design matrix was then directly regressed against the time-series from each \LW{parcel} for each subject, thereby obtaining the regional regression parameters.
The final values were obtained by averaging over all subjects.

\subsection{Statistical analysis}

\noindent
The difference in mean information content between the task fMRI time series with the highest information content and all other task fMRI time series was tested with a two-sided $t$-test.
The results were corrected with the Bonferroni method for multiple comparisons, as seven different tests were carried out~\cite{bonferroni_review}.
In the present study, the number of subjects was very large and the number of tests low, and thus the probability of type 2 errors was small. 
To show that information content contains information beyond the squared Euclidean distance, the predictiveness of information content together with squared Euclidean distance was compared to squared Euclidean distance only.
\LW{Because the data was available in pairs, the Wilcoxon signed-rank test was used for evaluating the significance of the difference in predictiveness.}
Further, the difference between the brain efficiency of low, middling, and high information content states was analyzed.
Here, the one-way ANOVA test was used. 
All statistical tests were implemented in the Scipy library~\cite{scipy}, with multiple comparison corrections taken from the statsmodels library~\cite{seabold2010statsmodels}.

\section*{Citation Diversity Statement}

\noindent
Recent work in several fields of science has identified a bias in citation practices such that papers from women and other minority scholars are under-cited relative to the number of such papers in the field \cite{mitchell2013gendered,dion2018gendered,caplar2017quantitative, maliniak2013gender, Dworkin20200103894378, bertolero2021racial, wang2021gendered, chatterjee2021gender, fulvio2021imbalance}. 
Here we sought to proactively consider choosing references that reflect the diversity of the field in thought, form of contribution, gender, race, ethnicity, and other factors. 
First, we obtained the predicted gender of the first and last author of each reference by using databases that store the probability of a first name being carried by a woman \cite{Dworkin20200103894378,zhou_dale_2020_3672110}. 
By this measure (and excluding self-citations to the first and last authors of our current paper), our references contain 14.97\% woman(first)/woman(last), 14.48\% man/woman, 13.58\% woman/man, and 56.97\% man/man. 
This method is limited in that a) names, pronouns, and social media profiles used to construct the databases may not, in every case, be indicative of gender identity and b) it cannot account for intersex, non-binary, or transgender people. 
Second, we obtained predicted racial/ethnic category of the first and last author of each reference by databases that store the probability of a first and last name being carried by an author of color \cite{ambekar2009name, sood2018predicting}. 
By this measure (and excluding self-citations), our references contain 14.41\% author of color (first)/author of color(last), 16.45\% white author/author of color, 22.37\% author of color/white author, and 46.77\% white author/white author. 
This method is limited in that a) names and Florida Voter Data to make the predictions may not be indicative of racial/ethnic identity, and b) it cannot account for Indigenous and mixed-race authors, or those who may face differential biases due to the ambiguous racialization or ethnicization of their names.  
We look forward to future work that could help us to better understand how to support equitable practices in science.

\section*{Funding}
\noindent
This work was funded by by the Deutsche Forschungsgemeinschaft (DFG, German Research Foundation) – grant 269953372/GRK2150.

\section*{Data availability statement}

\noindent
All data used in this study are openly available from the Human Connectome Project~\cite{hcp_overview}.
The code used for computing network control and IC as well as for performing the statistical analysis is available on Github (\url{https://github.com/weningerleon/InformationContent\textunderscore HCP}).

\clearpage

\bibliography{literature}

\begin{thebibliography}{92}%
\makeatletter
\providecommand \@ifxundefined [1]{%
 \@ifx{#1\undefined}
}%
\providecommand \@ifnum [1]{%
 \ifnum #1\expandafter \@firstoftwo
 \else \expandafter \@secondoftwo
 \fi
}%
\providecommand \@ifx [1]{%
 \ifx #1\expandafter \@firstoftwo
 \else \expandafter \@secondoftwo
 \fi
}%
\providecommand \natexlab [1]{#1}%
\providecommand \enquote  [1]{``#1''}%
\providecommand \bibnamefont  [1]{#1}%
\providecommand \bibfnamefont [1]{#1}%
\providecommand \citenamefont [1]{#1}%
\providecommand \href@noop [0]{\@secondoftwo}%
\providecommand \href [0]{\begingroup \@sanitize@url \@href}%
\providecommand \@href[1]{\@@startlink{#1}\@@href}%
\providecommand \@@href[1]{\endgroup#1\@@endlink}%
\providecommand \@sanitize@url [0]{\catcode `\\12\catcode `\$12\catcode
  `\&12\catcode `\#12\catcode `\^12\catcode `\_12\catcode `\%12\relax}%
\providecommand \@@startlink[1]{}%
\providecommand \@@endlink[0]{}%
\providecommand \url  [0]{\begingroup\@sanitize@url \@url }%
\providecommand \@url [1]{\endgroup\@href {#1}{\urlprefix }}%
\providecommand \urlprefix  [0]{URL }%
\providecommand \Eprint [0]{\href }%
\providecommand \doibase [0]{https://doi.org/}%
\providecommand \selectlanguage [0]{\@gobble}%
\providecommand \bibinfo  [0]{\@secondoftwo}%
\providecommand \bibfield  [0]{\@secondoftwo}%
\providecommand \translation [1]{[#1]}%
\providecommand \BibitemOpen [0]{}%
\providecommand \bibitemStop [0]{}%
\providecommand \bibitemNoStop [0]{.\EOS\space}%
\providecommand \EOS [0]{\spacefactor3000\relax}%
\providecommand \BibitemShut  [1]{\csname bibitem#1\endcsname}%
\let\auto@bib@innerbib\@empty
\bibitem [{\citenamefont {Avena-Koenigsberger}\ \emph
  {et~al.}(2018)\citenamefont {Avena-Koenigsberger}, \citenamefont {Misic},\
  and\ \citenamefont {Sporns}}]{avena2018communication}%
  \BibitemOpen
  \bibfield  {author} {\bibinfo {author} {\bibfnamefont {A.}~\bibnamefont
  {Avena-Koenigsberger}}, \bibinfo {author} {\bibfnamefont {B.}~\bibnamefont
  {Misic}},\ and\ \bibinfo {author} {\bibfnamefont {O.}~\bibnamefont
  {Sporns}},\ }\bibfield  {title} {\bibinfo {title} {Communication dynamics in
  complex brain networks},\ }\href@noop {} {\bibfield  {journal} {\bibinfo
  {journal} {Nature Reviews Neuroscience}\ }\textbf {\bibinfo {volume} {19}},\
  \bibinfo {pages} {17} (\bibinfo {year} {2018})}\BibitemShut {NoStop}%
\bibitem [{\citenamefont {Srivastava}\ \emph {et~al.}(2020)\citenamefont
  {Srivastava}, \citenamefont {Nozari}, \citenamefont {Kim}, \citenamefont
  {Ju}, \citenamefont {Zhou}, \citenamefont {Becker}, \citenamefont
  {Pasqualetti}, \citenamefont {Pappas},\ and\ \citenamefont
  {Bassett}}]{srivastava2020models}%
  \BibitemOpen
  \bibfield  {author} {\bibinfo {author} {\bibfnamefont {P.}~\bibnamefont
  {Srivastava}}, \bibinfo {author} {\bibfnamefont {E.}~\bibnamefont {Nozari}},
  \bibinfo {author} {\bibfnamefont {J.~Z.}\ \bibnamefont {Kim}}, \bibinfo
  {author} {\bibfnamefont {H.}~\bibnamefont {Ju}}, \bibinfo {author}
  {\bibfnamefont {D.}~\bibnamefont {Zhou}}, \bibinfo {author} {\bibfnamefont
  {C.}~\bibnamefont {Becker}}, \bibinfo {author} {\bibfnamefont
  {F.}~\bibnamefont {Pasqualetti}}, \bibinfo {author} {\bibfnamefont {G.~J.}\
  \bibnamefont {Pappas}},\ and\ \bibinfo {author} {\bibfnamefont {D.~S.}\
  \bibnamefont {Bassett}},\ }\bibfield  {title} {\bibinfo {title} {Models of
  communication and control for brain networks: distinctions, convergence, and
  future outlook},\ }\href@noop {} {\bibfield  {journal} {\bibinfo  {journal}
  {Netw Neurosci}\ }\textbf {\bibinfo {volume} {4}},\ \bibinfo {pages} {1122}
  (\bibinfo {year} {2020})}\BibitemShut {NoStop}%
\bibitem [{\citenamefont {Jbabdi}\ and\ \citenamefont
  {Johansen-Berg}(2011)}]{jbabdi2011tractography}%
  \BibitemOpen
  \bibfield  {author} {\bibinfo {author} {\bibfnamefont {S.}~\bibnamefont
  {Jbabdi}}\ and\ \bibinfo {author} {\bibfnamefont {H.}~\bibnamefont
  {Johansen-Berg}},\ }\bibfield  {title} {\bibinfo {title} {Tractography: where
  do we go from here?},\ }\href@noop {} {\bibfield  {journal} {\bibinfo
  {journal} {Brain Connect}\ }\textbf {\bibinfo {volume} {1}},\ \bibinfo
  {pages} {169} (\bibinfo {year} {2011})}\BibitemShut {NoStop}%
\bibitem [{\citenamefont {Johansen-Berg}(2013)}]{johansenberg2013human}%
  \BibitemOpen
  \bibfield  {author} {\bibinfo {author} {\bibfnamefont {H.}~\bibnamefont
  {Johansen-Berg}},\ }\bibfield  {title} {\bibinfo {title} {Human connectomics
  - what will the future demand?},\ }\href@noop {} {\bibfield  {journal}
  {\bibinfo  {journal} {Neuroimage}\ }\textbf {\bibinfo {volume} {80}},\
  \bibinfo {pages} {541} (\bibinfo {year} {2013})}\BibitemShut {NoStop}%
\bibitem [{\citenamefont {Sorrentino}\ \emph {et~al.}(2021)\citenamefont
  {Sorrentino}, \citenamefont {Seguin}, \citenamefont {Rucco}, \citenamefont
  {Liparoti}, \citenamefont {Lopez}, \citenamefont {Bonavita}, \citenamefont
  {Quarantelli}, \citenamefont {Sorrentino}, \citenamefont {Jirsa},\ and\
  \citenamefont {Zalesky}}]{sorrentino2021structural}%
  \BibitemOpen
  \bibfield  {author} {\bibinfo {author} {\bibfnamefont {P.}~\bibnamefont
  {Sorrentino}}, \bibinfo {author} {\bibfnamefont {C.}~\bibnamefont {Seguin}},
  \bibinfo {author} {\bibfnamefont {R.}~\bibnamefont {Rucco}}, \bibinfo
  {author} {\bibfnamefont {M.}~\bibnamefont {Liparoti}}, \bibinfo {author}
  {\bibfnamefont {E.~T.}\ \bibnamefont {Lopez}}, \bibinfo {author}
  {\bibfnamefont {S.}~\bibnamefont {Bonavita}}, \bibinfo {author}
  {\bibfnamefont {M.}~\bibnamefont {Quarantelli}}, \bibinfo {author}
  {\bibfnamefont {G.}~\bibnamefont {Sorrentino}}, \bibinfo {author}
  {\bibfnamefont {V.}~\bibnamefont {Jirsa}},\ and\ \bibinfo {author}
  {\bibfnamefont {A.}~\bibnamefont {Zalesky}},\ }\bibfield  {title} {\bibinfo
  {title} {The structural connectome constrains fast brain dynamics},\
  }\href@noop {} {\bibfield  {journal} {\bibinfo  {journal} {Elife}\ }\textbf
  {\bibinfo {volume} {10}} (\bibinfo {year} {2021})}\BibitemShut {NoStop}%
\bibitem [{\citenamefont {Imms}\ \emph {et~al.}(2021)\citenamefont {Imms},
  \citenamefont {Domínguez~D}, \citenamefont {Burmester}, \citenamefont
  {Seguin}, \citenamefont {Clemente}, \citenamefont {Dhollander}, \citenamefont
  {Wilson}, \citenamefont {Poudel},\ and\ \citenamefont
  {Caeyenberghs}}]{imms2021navigating}%
  \BibitemOpen
  \bibfield  {author} {\bibinfo {author} {\bibfnamefont {P.}~\bibnamefont
  {Imms}}, \bibinfo {author} {\bibfnamefont {J.~F.}\ \bibnamefont
  {Domínguez~D}}, \bibinfo {author} {\bibfnamefont {A.}~\bibnamefont
  {Burmester}}, \bibinfo {author} {\bibfnamefont {C.}~\bibnamefont {Seguin}},
  \bibinfo {author} {\bibfnamefont {A.}~\bibnamefont {Clemente}}, \bibinfo
  {author} {\bibfnamefont {T.}~\bibnamefont {Dhollander}}, \bibinfo {author}
  {\bibfnamefont {P.~H.}\ \bibnamefont {Wilson}}, \bibinfo {author}
  {\bibfnamefont {G.}~\bibnamefont {Poudel}},\ and\ \bibinfo {author}
  {\bibfnamefont {K.}~\bibnamefont {Caeyenberghs}},\ }\bibfield  {title}
  {\bibinfo {title} {Navigating the link between processing speed and network
  communication in the human brain},\ }\href@noop {} {\bibfield  {journal}
  {\bibinfo  {journal} {Brain Struct Funct}\ }\textbf {\bibinfo {volume}
  {226}},\ \bibinfo {pages} {1281} (\bibinfo {year} {2021})}\BibitemShut
  {NoStop}%
\bibitem [{\citenamefont {Ju}\ and\ \citenamefont
  {Bassett}(2020)}]{ju2020dynamic}%
  \BibitemOpen
  \bibfield  {author} {\bibinfo {author} {\bibfnamefont {H.}~\bibnamefont
  {Ju}}\ and\ \bibinfo {author} {\bibfnamefont {D.~S.}\ \bibnamefont
  {Bassett}},\ }\bibfield  {title} {\bibinfo {title} {Dynamic representations
  in networked neural systems},\ }\href@noop {} {\bibfield  {journal} {\bibinfo
   {journal} {Nat Neurosci}\ }\textbf {\bibinfo {volume} {23}},\ \bibinfo
  {pages} {908} (\bibinfo {year} {2020})}\BibitemShut {NoStop}%
\bibitem [{\citenamefont {Beynel}\ \emph {et~al.}(2020)\citenamefont {Beynel},
  \citenamefont {Deng}, \citenamefont {Crowell}, \citenamefont {Hilbig},
  \citenamefont {Peterchev}, \citenamefont {Luber}, \citenamefont {Lisanby},
  \citenamefont {Cabeza}, \citenamefont {Appelbaum},\ and\ \citenamefont
  {Davis}}]{beynel2020structural}%
  \BibitemOpen
  \bibfield  {author} {\bibinfo {author} {\bibfnamefont {L.}~\bibnamefont
  {Beynel}}, \bibinfo {author} {\bibfnamefont {L.}~\bibnamefont {Deng}},
  \bibinfo {author} {\bibfnamefont {C.~A.}\ \bibnamefont {Crowell}}, \bibinfo
  {author} {\bibfnamefont {S.}~\bibnamefont {Hilbig}}, \bibinfo {author}
  {\bibfnamefont {A.~V.}\ \bibnamefont {Peterchev}}, \bibinfo {author}
  {\bibfnamefont {B.}~\bibnamefont {Luber}}, \bibinfo {author} {\bibfnamefont
  {S.~H.}\ \bibnamefont {Lisanby}}, \bibinfo {author} {\bibfnamefont
  {R.}~\bibnamefont {Cabeza}}, \bibinfo {author} {\bibfnamefont {L.~G.}\
  \bibnamefont {Appelbaum}},\ and\ \bibinfo {author} {\bibfnamefont {S.~W.}\
  \bibnamefont {Davis}},\ }\bibfield  {title} {\bibinfo {title} {Structural
  controllability predicts functional patterns and brain stimulation benefits
  associated with working memory},\ }\href@noop {} {\bibfield  {journal}
  {\bibinfo  {journal} {J Neurosci}\ }\textbf {\bibinfo {volume} {40}},\
  \bibinfo {pages} {6770} (\bibinfo {year} {2020})}\BibitemShut {NoStop}%
\bibitem [{\citenamefont {Towlson}\ and\ \citenamefont
  {Barabási}(2020)}]{towlson2020synthetic}%
  \BibitemOpen
  \bibfield  {author} {\bibinfo {author} {\bibfnamefont {E.~K.}\ \bibnamefont
  {Towlson}}\ and\ \bibinfo {author} {\bibfnamefont {A.-L.}\ \bibnamefont
  {Barabási}},\ }\bibfield  {title} {\bibinfo {title} {Synthetic ablations in
  the {C.} elegans nervous system},\ }\href@noop {} {\bibfield  {journal}
  {\bibinfo  {journal} {Netw Neurosci}\ }\textbf {\bibinfo {volume} {4}},\
  \bibinfo {pages} {200} (\bibinfo {year} {2020})}\BibitemShut {NoStop}%
\bibitem [{\citenamefont {Singleton}\ \emph {et~al.}(2021)\citenamefont
  {Singleton}, \citenamefont {Luppi}, \citenamefont {Carhart-Harris},
  \citenamefont {Cruzat}, \citenamefont {Roseman}, \citenamefont {Deco},
  \citenamefont {Kringelbach}, \citenamefont {Stamatakis},\ and\ \citenamefont
  {Kuceyeski}}]{singleton2021lsd}%
  \BibitemOpen
  \bibfield  {author} {\bibinfo {author} {\bibfnamefont {S.~P.}\ \bibnamefont
  {Singleton}}, \bibinfo {author} {\bibfnamefont {A.~I.}\ \bibnamefont
  {Luppi}}, \bibinfo {author} {\bibfnamefont {R.~L.}\ \bibnamefont
  {Carhart-Harris}}, \bibinfo {author} {\bibfnamefont {J.}~\bibnamefont
  {Cruzat}}, \bibinfo {author} {\bibfnamefont {L.}~\bibnamefont {Roseman}},
  \bibinfo {author} {\bibfnamefont {G.}~\bibnamefont {Deco}}, \bibinfo {author}
  {\bibfnamefont {M.~L.}\ \bibnamefont {Kringelbach}}, \bibinfo {author}
  {\bibfnamefont {E.~A.}\ \bibnamefont {Stamatakis}},\ and\ \bibinfo {author}
  {\bibfnamefont {A.}~\bibnamefont {Kuceyeski}},\ }\bibfield  {title} {\bibinfo
  {title} {{LSD} flattens the brain’s energy landscape: evidence from
  receptor-informed network control theory},\ }\href@noop {} {\bibfield
  {journal} {\bibinfo  {journal} {bioRxiv}\ } (\bibinfo {year}
  {2021})}\BibitemShut {NoStop}%
\bibitem [{\citenamefont {Gu}\ \emph {et~al.}(2015)\citenamefont {Gu},
  \citenamefont {Pasqualetti}, \citenamefont {Cieslak}, \citenamefont
  {Telesford}, \citenamefont {Alfred}, \citenamefont {Kahn}, \citenamefont
  {Medaglia}, \citenamefont {Vettel}, \citenamefont {Miller}, \citenamefont
  {Grafton} \emph {et~al.}}]{gu2015controllability}%
  \BibitemOpen
  \bibfield  {author} {\bibinfo {author} {\bibfnamefont {S.}~\bibnamefont
  {Gu}}, \bibinfo {author} {\bibfnamefont {F.}~\bibnamefont {Pasqualetti}},
  \bibinfo {author} {\bibfnamefont {M.}~\bibnamefont {Cieslak}}, \bibinfo
  {author} {\bibfnamefont {Q.~K.}\ \bibnamefont {Telesford}}, \bibinfo {author}
  {\bibfnamefont {B.~Y.}\ \bibnamefont {Alfred}}, \bibinfo {author}
  {\bibfnamefont {A.~E.}\ \bibnamefont {Kahn}}, \bibinfo {author}
  {\bibfnamefont {J.~D.}\ \bibnamefont {Medaglia}}, \bibinfo {author}
  {\bibfnamefont {J.~M.}\ \bibnamefont {Vettel}}, \bibinfo {author}
  {\bibfnamefont {M.~B.}\ \bibnamefont {Miller}}, \bibinfo {author}
  {\bibfnamefont {S.~T.}\ \bibnamefont {Grafton}}, \emph {et~al.},\ }\bibfield
  {title} {\bibinfo {title} {Controllability of structural brain networks},\
  }\href@noop {} {\bibfield  {journal} {\bibinfo  {journal} {Nature
  communications}\ }\textbf {\bibinfo {volume} {6}},\ \bibinfo {pages} {1}
  (\bibinfo {year} {2015})}\BibitemShut {NoStop}%
\bibitem [{\citenamefont {Pasqualetti}\ \emph {et~al.}(2014)\citenamefont
  {Pasqualetti}, \citenamefont {Zampieri},\ and\ \citenamefont
  {Bullo}}]{pasqualetti2014controllability}%
  \BibitemOpen
  \bibfield  {author} {\bibinfo {author} {\bibfnamefont {F.}~\bibnamefont
  {Pasqualetti}}, \bibinfo {author} {\bibfnamefont {S.}~\bibnamefont
  {Zampieri}},\ and\ \bibinfo {author} {\bibfnamefont {F.}~\bibnamefont
  {Bullo}},\ }\bibfield  {title} {\bibinfo {title} {Controllability metrics,
  limitations and algorithms for complex networks},\ }in\ \href
  {https://doi.org/10.1109/ACC.2014.6858621} {\emph {\bibinfo {booktitle} {2014
  American Control Conference}}}\ (\bibinfo {year} {2014})\ pp.\ \bibinfo
  {pages} {3287--3292}\BibitemShut {NoStop}%
\bibitem [{\citenamefont {Liu}\ \emph {et~al.}(2011)\citenamefont {Liu},
  \citenamefont {Slotine},\ and\ \citenamefont
  {Barabási}}]{liu2011controllability}%
  \BibitemOpen
  \bibfield  {author} {\bibinfo {author} {\bibfnamefont {Y.-Y.}\ \bibnamefont
  {Liu}}, \bibinfo {author} {\bibfnamefont {J.-J.}\ \bibnamefont {Slotine}},\
  and\ \bibinfo {author} {\bibfnamefont {A.-L.}\ \bibnamefont {Barabási}},\
  }\bibfield  {title} {\bibinfo {title} {Controllability of complex networks},\
  }\href@noop {} {\bibfield  {journal} {\bibinfo  {journal} {Nature}\ }\textbf
  {\bibinfo {volume} {473}},\ \bibinfo {pages} {167} (\bibinfo {year}
  {2011})}\BibitemShut {NoStop}%
\bibitem [{\citenamefont {Betzel}\ \emph {et~al.}(2016)\citenamefont {Betzel},
  \citenamefont {Gu}, \citenamefont {Medaglia}, \citenamefont {Pasqualetti},\
  and\ \citenamefont {Bassett}}]{betzel2016optimally}%
  \BibitemOpen
  \bibfield  {author} {\bibinfo {author} {\bibfnamefont {R.~F.}\ \bibnamefont
  {Betzel}}, \bibinfo {author} {\bibfnamefont {S.}~\bibnamefont {Gu}}, \bibinfo
  {author} {\bibfnamefont {J.~D.}\ \bibnamefont {Medaglia}}, \bibinfo {author}
  {\bibfnamefont {F.}~\bibnamefont {Pasqualetti}},\ and\ \bibinfo {author}
  {\bibfnamefont {D.~S.}\ \bibnamefont {Bassett}},\ }\bibfield  {title}
  {\bibinfo {title} {Optimally controlling the human connectome: the role of
  network topology},\ }\href@noop {} {\bibfield  {journal} {\bibinfo  {journal}
  {Sci Rep}\ }\textbf {\bibinfo {volume} {6}},\ \bibinfo {pages} {30770}
  (\bibinfo {year} {2016})}\BibitemShut {NoStop}%
\bibitem [{\citenamefont {Gu}\ \emph {et~al.}(2017)\citenamefont {Gu},
  \citenamefont {Betzel}, \citenamefont {Mattar}, \citenamefont {Cieslak},
  \citenamefont {Delio}, \citenamefont {Grafton}, \citenamefont {Pasqualetti},\
  and\ \citenamefont {Bassett}}]{gu2017optimal}%
  \BibitemOpen
  \bibfield  {author} {\bibinfo {author} {\bibfnamefont {S.}~\bibnamefont
  {Gu}}, \bibinfo {author} {\bibfnamefont {R.~F.}\ \bibnamefont {Betzel}},
  \bibinfo {author} {\bibfnamefont {M.~G.}\ \bibnamefont {Mattar}}, \bibinfo
  {author} {\bibfnamefont {M.}~\bibnamefont {Cieslak}}, \bibinfo {author}
  {\bibfnamefont {P.~R.}\ \bibnamefont {Delio}}, \bibinfo {author}
  {\bibfnamefont {S.~T.}\ \bibnamefont {Grafton}}, \bibinfo {author}
  {\bibfnamefont {F.}~\bibnamefont {Pasqualetti}},\ and\ \bibinfo {author}
  {\bibfnamefont {D.~S.}\ \bibnamefont {Bassett}},\ }\bibfield  {title}
  {\bibinfo {title} {Optimal trajectories of brain state transitions},\
  }\href@noop {} {\bibfield  {journal} {\bibinfo  {journal} {Neuroimage}\
  }\textbf {\bibinfo {volume} {148}},\ \bibinfo {pages} {305} (\bibinfo {year}
  {2017})}\BibitemShut {NoStop}%
\bibitem [{\citenamefont {Shannon}(1948)}]{shannon_info}%
  \BibitemOpen
  \bibfield  {author} {\bibinfo {author} {\bibfnamefont {C.}~\bibnamefont
  {Shannon}},\ }\bibfield  {title} {\bibinfo {title} {A mathematical theory of
  communication},\ }\href {https://doi.org/10.1002/j.1538-7305.1948.tb01338.x}
  {\bibfield  {journal} {\bibinfo  {journal} {The Bell System Technical
  Journal}\ }\textbf {\bibinfo {volume} {27}},\ \bibinfo {pages} {379}
  (\bibinfo {year} {1948})}\BibitemShut {NoStop}%
\bibitem [{\citenamefont {Collell}\ and\ \citenamefont
  {Fauquet}(2015)}]{collell2015brain}%
  \BibitemOpen
  \bibfield  {author} {\bibinfo {author} {\bibfnamefont {G.}~\bibnamefont
  {Collell}}\ and\ \bibinfo {author} {\bibfnamefont {J.}~\bibnamefont
  {Fauquet}},\ }\bibfield  {title} {\bibinfo {title} {Brain activity and
  cognition: a connection from thermodynamics and information theory},\
  }\href@noop {} {\bibfield  {journal} {\bibinfo  {journal} {Frontiers in
  psychology}\ }\textbf {\bibinfo {volume} {6}},\ \bibinfo {pages} {818}
  (\bibinfo {year} {2015})}\BibitemShut {NoStop}%
\bibitem [{\citenamefont {Barch}\ \emph
  {et~al.}(2013{\natexlab{a}})\citenamefont {Barch}, \citenamefont {Burgess},
  \citenamefont {Harms}, \citenamefont {Petersen}, \citenamefont {Schlaggar},
  \citenamefont {Corbetta}, \citenamefont {Glasser}, \citenamefont {Curtiss},
  \citenamefont {Dixit}, \citenamefont {Feldt}, \citenamefont {Nolan},
  \citenamefont {Bryant}, \citenamefont {Hartley}, \citenamefont {Footer},
  \citenamefont {Bjork}, \citenamefont {Poldrack}, \citenamefont {Smith},
  \citenamefont {Johansen-Berg}, \citenamefont {Snyder}, \citenamefont
  {Van~Essen},\ and\ \citenamefont {Consortium}}]{barch2013function}%
  \BibitemOpen
  \bibfield  {author} {\bibinfo {author} {\bibfnamefont {D.~M.}\ \bibnamefont
  {Barch}}, \bibinfo {author} {\bibfnamefont {G.~C.}\ \bibnamefont {Burgess}},
  \bibinfo {author} {\bibfnamefont {M.~P.}\ \bibnamefont {Harms}}, \bibinfo
  {author} {\bibfnamefont {S.~E.}\ \bibnamefont {Petersen}}, \bibinfo {author}
  {\bibfnamefont {B.~L.}\ \bibnamefont {Schlaggar}}, \bibinfo {author}
  {\bibfnamefont {M.}~\bibnamefont {Corbetta}}, \bibinfo {author}
  {\bibfnamefont {M.~F.}\ \bibnamefont {Glasser}}, \bibinfo {author}
  {\bibfnamefont {S.}~\bibnamefont {Curtiss}}, \bibinfo {author} {\bibfnamefont
  {S.}~\bibnamefont {Dixit}}, \bibinfo {author} {\bibfnamefont
  {C.}~\bibnamefont {Feldt}}, \bibinfo {author} {\bibfnamefont
  {D.}~\bibnamefont {Nolan}}, \bibinfo {author} {\bibfnamefont
  {E.}~\bibnamefont {Bryant}}, \bibinfo {author} {\bibfnamefont
  {T.}~\bibnamefont {Hartley}}, \bibinfo {author} {\bibfnamefont
  {O.}~\bibnamefont {Footer}}, \bibinfo {author} {\bibfnamefont {J.~M.}\
  \bibnamefont {Bjork}}, \bibinfo {author} {\bibfnamefont {R.}~\bibnamefont
  {Poldrack}}, \bibinfo {author} {\bibfnamefont {S.}~\bibnamefont {Smith}},
  \bibinfo {author} {\bibfnamefont {H.}~\bibnamefont {Johansen-Berg}}, \bibinfo
  {author} {\bibfnamefont {A.~Z.}\ \bibnamefont {Snyder}}, \bibinfo {author}
  {\bibfnamefont {D.~C.}\ \bibnamefont {Van~Essen}},\ and\ \bibinfo {author}
  {\bibfnamefont {W.-M.~H.}\ \bibnamefont {Consortium}},\ }\bibfield  {title}
  {\bibinfo {title} {Function in the human connectome: task-{fMRI} and
  individual differences in behavior},\ }\href@noop {} {\bibfield  {journal}
  {\bibinfo  {journal} {Neuroimage}\ }\textbf {\bibinfo {volume} {80}},\
  \bibinfo {pages} {169} (\bibinfo {year} {2013}{\natexlab{a}})}\BibitemShut
  {NoStop}%
\bibitem [{\citenamefont {Lynn}\ \emph {et~al.}(2020)\citenamefont {Lynn},
  \citenamefont {Cornblath}, \citenamefont {Papadopoulos}, \citenamefont
  {Bertolero},\ and\ \citenamefont {Bassett}}]{lynn2021broken}%
  \BibitemOpen
  \bibfield  {author} {\bibinfo {author} {\bibfnamefont {C.~W.}\ \bibnamefont
  {Lynn}}, \bibinfo {author} {\bibfnamefont {E.~J.}\ \bibnamefont {Cornblath}},
  \bibinfo {author} {\bibfnamefont {L.}~\bibnamefont {Papadopoulos}}, \bibinfo
  {author} {\bibfnamefont {M.~A.}\ \bibnamefont {Bertolero}},\ and\ \bibinfo
  {author} {\bibfnamefont {D.~S.}\ \bibnamefont {Bassett}},\ }\bibfield
  {title} {\bibinfo {title} {Broken detailed balance and entropy production in
  the human brain},\ }\href@noop {} {\bibfield  {journal} {\bibinfo  {journal}
  {arXiv}\ }\textbf {\bibinfo {volume} {2005}},\ \bibinfo {pages} {02526}
  (\bibinfo {year} {2020})}\BibitemShut {NoStop}%
\bibitem [{\citenamefont {Gonzalez-Castillo}\ \emph {et~al.}(2021)\citenamefont
  {Gonzalez-Castillo}, \citenamefont {Kam}, \citenamefont {Hoy},\ and\
  \citenamefont {Bandettini}}]{gonzalez2021how}%
  \BibitemOpen
  \bibfield  {author} {\bibinfo {author} {\bibfnamefont {J.}~\bibnamefont
  {Gonzalez-Castillo}}, \bibinfo {author} {\bibfnamefont {J.~W.~Y.}\
  \bibnamefont {Kam}}, \bibinfo {author} {\bibfnamefont {C.~W.}\ \bibnamefont
  {Hoy}},\ and\ \bibinfo {author} {\bibfnamefont {P.~A.}\ \bibnamefont
  {Bandettini}},\ }\bibfield  {title} {\bibinfo {title} {How to interpret
  resting-state {fMRI}: Ask your participants},\ }\href@noop {} {\bibfield
  {journal} {\bibinfo  {journal} {J Neurosci}\ }\textbf {\bibinfo {volume}
  {41}},\ \bibinfo {pages} {1130} (\bibinfo {year} {2021})}\BibitemShut
  {NoStop}%
\bibitem [{\citenamefont {Cohen}(2018)}]{cohen2018behavioral}%
  \BibitemOpen
  \bibfield  {author} {\bibinfo {author} {\bibfnamefont {J.~R.}\ \bibnamefont
  {Cohen}},\ }\bibfield  {title} {\bibinfo {title} {The behavioral and
  cognitive relevance of time-varying, dynamic changes in functional
  connectivity},\ }\href@noop {} {\bibfield  {journal} {\bibinfo  {journal}
  {Neuroimage}\ }\textbf {\bibinfo {volume} {180}},\ \bibinfo {pages} {515}
  (\bibinfo {year} {2018})}\BibitemShut {NoStop}%
\bibitem [{\citenamefont {Vukovic}\ \emph {et~al.}(2021)\citenamefont
  {Vukovic}, \citenamefont {Hansen}, \citenamefont {Lund}, \citenamefont
  {Jespersen},\ and\ \citenamefont {Shtyrov}}]{vukovic2021rapid}%
  \BibitemOpen
  \bibfield  {author} {\bibinfo {author} {\bibfnamefont {N.}~\bibnamefont
  {Vukovic}}, \bibinfo {author} {\bibfnamefont {B.}~\bibnamefont {Hansen}},
  \bibinfo {author} {\bibfnamefont {T.~E.}\ \bibnamefont {Lund}}, \bibinfo
  {author} {\bibfnamefont {S.}~\bibnamefont {Jespersen}},\ and\ \bibinfo
  {author} {\bibfnamefont {Y.}~\bibnamefont {Shtyrov}},\ }\bibfield  {title}
  {\bibinfo {title} {Rapid microstructural plasticity in the cortical semantic
  network following a short language learning session},\ }\href@noop {}
  {\bibfield  {journal} {\bibinfo  {journal} {PLoS Biol}\ }\textbf {\bibinfo
  {volume} {19}},\ \bibinfo {pages} {e3001290} (\bibinfo {year}
  {2021})}\BibitemShut {NoStop}%
\bibitem [{\citenamefont {Sagi}\ \emph {et~al.}(2012)\citenamefont {Sagi},
  \citenamefont {Tavor}, \citenamefont {Hofstetter}, \citenamefont
  {Tzur-Moryosef}, \citenamefont {Blumenfeld-Katzir},\ and\ \citenamefont
  {Assaf}}]{sagi2012learning}%
  \BibitemOpen
  \bibfield  {author} {\bibinfo {author} {\bibfnamefont {Y.}~\bibnamefont
  {Sagi}}, \bibinfo {author} {\bibfnamefont {I.}~\bibnamefont {Tavor}},
  \bibinfo {author} {\bibfnamefont {S.}~\bibnamefont {Hofstetter}}, \bibinfo
  {author} {\bibfnamefont {S.}~\bibnamefont {Tzur-Moryosef}}, \bibinfo {author}
  {\bibfnamefont {T.}~\bibnamefont {Blumenfeld-Katzir}},\ and\ \bibinfo
  {author} {\bibfnamefont {Y.}~\bibnamefont {Assaf}},\ }\bibfield  {title}
  {\bibinfo {title} {Learning in the fast lane: new insights into
  neuroplasticity},\ }\href@noop {} {\bibfield  {journal} {\bibinfo  {journal}
  {Neuron}\ }\textbf {\bibinfo {volume} {73}},\ \bibinfo {pages} {1195}
  (\bibinfo {year} {2012})}\BibitemShut {NoStop}%
\bibitem [{\citenamefont {Ilg}\ \emph {et~al.}(2008)\citenamefont {Ilg},
  \citenamefont {Wohlschläger}, \citenamefont {Gaser}, \citenamefont {Liebau},
  \citenamefont {Dauner}, \citenamefont {Wöller}, \citenamefont {Zimmer},
  \citenamefont {Zihl},\ and\ \citenamefont {Mühlau}}]{ilg2008gray}%
  \BibitemOpen
  \bibfield  {author} {\bibinfo {author} {\bibfnamefont {R.}~\bibnamefont
  {Ilg}}, \bibinfo {author} {\bibfnamefont {A.~M.}\ \bibnamefont
  {Wohlschläger}}, \bibinfo {author} {\bibfnamefont {C.}~\bibnamefont
  {Gaser}}, \bibinfo {author} {\bibfnamefont {Y.}~\bibnamefont {Liebau}},
  \bibinfo {author} {\bibfnamefont {R.}~\bibnamefont {Dauner}}, \bibinfo
  {author} {\bibfnamefont {A.}~\bibnamefont {Wöller}}, \bibinfo {author}
  {\bibfnamefont {C.}~\bibnamefont {Zimmer}}, \bibinfo {author} {\bibfnamefont
  {J.}~\bibnamefont {Zihl}},\ and\ \bibinfo {author} {\bibfnamefont
  {M.}~\bibnamefont {Mühlau}},\ }\bibfield  {title} {\bibinfo {title} {Gray
  matter increase induced by practice correlates with task-specific activation:
  a combined functional and morphometric magnetic resonance imaging study},\
  }\href@noop {} {\bibfield  {journal} {\bibinfo  {journal} {J Neurosci}\
  }\textbf {\bibinfo {volume} {28}},\ \bibinfo {pages} {4210} (\bibinfo {year}
  {2008})}\BibitemShut {NoStop}%
\bibitem [{\citenamefont {Wendelken}\ \emph {et~al.}(2017)\citenamefont
  {Wendelken}, \citenamefont {Ferrer}, \citenamefont {Ghetti}, \citenamefont
  {Bailey}, \citenamefont {Cutting},\ and\ \citenamefont
  {Bunge}}]{wendelken2017frontoparietal}%
  \BibitemOpen
  \bibfield  {author} {\bibinfo {author} {\bibfnamefont {C.}~\bibnamefont
  {Wendelken}}, \bibinfo {author} {\bibfnamefont {E.}~\bibnamefont {Ferrer}},
  \bibinfo {author} {\bibfnamefont {S.}~\bibnamefont {Ghetti}}, \bibinfo
  {author} {\bibfnamefont {S.~K.}\ \bibnamefont {Bailey}}, \bibinfo {author}
  {\bibfnamefont {L.}~\bibnamefont {Cutting}},\ and\ \bibinfo {author}
  {\bibfnamefont {S.~A.}\ \bibnamefont {Bunge}},\ }\bibfield  {title} {\bibinfo
  {title} {Frontoparietal structural connectivity in childhood predicts
  development of functional connectivity and reasoning ability: {A} large-scale
  longitudinal investigation},\ }\href@noop {} {\bibfield  {journal} {\bibinfo
  {journal} {J Neurosci}\ }\textbf {\bibinfo {volume} {37}},\ \bibinfo {pages}
  {8549} (\bibinfo {year} {2017})}\BibitemShut {NoStop}%
\bibitem [{\citenamefont {Raichle}\ and\ \citenamefont
  {Gusnard}(2002)}]{raichle2002appraising}%
  \BibitemOpen
  \bibfield  {author} {\bibinfo {author} {\bibfnamefont {M.~E.}\ \bibnamefont
  {Raichle}}\ and\ \bibinfo {author} {\bibfnamefont {D.~A.}\ \bibnamefont
  {Gusnard}},\ }\bibfield  {title} {\bibinfo {title} {Appraising the brain's
  energy budget},\ }\href@noop {} {\bibfield  {journal} {\bibinfo  {journal}
  {Proc. Natl. Acad. Sci.}\ }\textbf {\bibinfo {volume} {99}},\ \bibinfo
  {pages} {10237} (\bibinfo {year} {2002})}\BibitemShut {NoStop}%
\bibitem [{\citenamefont {Borda}(2011)}]{borda2011fundamentals}%
  \BibitemOpen
  \bibfield  {author} {\bibinfo {author} {\bibfnamefont {M.}~\bibnamefont
  {Borda}},\ }\href@noop {} {\emph {\bibinfo {title} {Fundamentals in
  Information Theory and Coding}}}\ (\bibinfo  {publisher} {Springer Berlin
  Heidelberg},\ \bibinfo {year} {2011})\BibitemShut {NoStop}%
\bibitem [{\citenamefont {Bishop}(2006)}]{bishopbook}%
  \BibitemOpen
  \bibfield  {author} {\bibinfo {author} {\bibfnamefont {C.~M.}\ \bibnamefont
  {Bishop}},\ }\href@noop {} {\emph {\bibinfo {title} {Pattern Recognition and
  Machine Learning (Information Science and Statistics)}}}\ (\bibinfo
  {publisher} {Springer-Verlag},\ \bibinfo {address} {Berlin, Heidelberg},\
  \bibinfo {year} {2006})\BibitemShut {NoStop}%
\bibitem [{\citenamefont {Schaefer}\ \emph {et~al.}(2017)\citenamefont
  {Schaefer}, \citenamefont {Kong}, \citenamefont {Gordon}, \citenamefont
  {Laumann}, \citenamefont {Zuo}, \citenamefont {Holmes}, \citenamefont
  {Eickhoff},\ and\ \citenamefont {Yeo}}]{schaefer_atlas}%
  \BibitemOpen
  \bibfield  {author} {\bibinfo {author} {\bibfnamefont {A.}~\bibnamefont
  {Schaefer}}, \bibinfo {author} {\bibfnamefont {R.}~\bibnamefont {Kong}},
  \bibinfo {author} {\bibfnamefont {E.~M.}\ \bibnamefont {Gordon}}, \bibinfo
  {author} {\bibfnamefont {T.~O.}\ \bibnamefont {Laumann}}, \bibinfo {author}
  {\bibfnamefont {X.-N.}\ \bibnamefont {Zuo}}, \bibinfo {author} {\bibfnamefont
  {A.~J.}\ \bibnamefont {Holmes}}, \bibinfo {author} {\bibfnamefont {S.~B.}\
  \bibnamefont {Eickhoff}},\ and\ \bibinfo {author} {\bibfnamefont {B.~T.~T.}\
  \bibnamefont {Yeo}},\ }\bibfield  {title} {\bibinfo {title} {{Local-Global
  Parcellation of the Human Cerebral Cortex from Intrinsic Functional
  Connectivity MRI}},\ }\href {https://doi.org/10.1093/cercor/bhx179}
  {\bibfield  {journal} {\bibinfo  {journal} {Cerebral Cortex}\ }\textbf
  {\bibinfo {volume} {28}},\ \bibinfo {pages} {3095} (\bibinfo {year}
  {2017})}\BibitemShut {NoStop}%
\bibitem [{\citenamefont {Tsai}\ \emph {et~al.}(1999)\citenamefont {Tsai},
  \citenamefont {Fisher}, \citenamefont {Wible}, \citenamefont {Wells},
  \citenamefont {Kim},\ and\ \citenamefont {Willsky}}]{mi_fmri}%
  \BibitemOpen
  \bibfield  {author} {\bibinfo {author} {\bibfnamefont {A.}~\bibnamefont
  {Tsai}}, \bibinfo {author} {\bibfnamefont {J.~W.}\ \bibnamefont {Fisher}},
  \bibinfo {author} {\bibfnamefont {C.}~\bibnamefont {Wible}}, \bibinfo
  {author} {\bibfnamefont {W.~M.}\ \bibnamefont {Wells}}, \bibinfo {author}
  {\bibfnamefont {J.}~\bibnamefont {Kim}},\ and\ \bibinfo {author}
  {\bibfnamefont {A.~S.}\ \bibnamefont {Willsky}},\ }\bibfield  {title}
  {\bibinfo {title} {Analysis of functional mri data using mutual
  information},\ }in\ \href@noop {} {\emph {\bibinfo {booktitle} {Medical Image
  Computing and Computer-Assisted Intervention -- MICCAI'99}}},\ \bibinfo
  {editor} {edited by\ \bibinfo {editor} {\bibfnamefont {C.}~\bibnamefont
  {Taylor}}\ and\ \bibinfo {editor} {\bibfnamefont {A.}~\bibnamefont
  {Colchester}}}\ (\bibinfo  {publisher} {Springer Berlin Heidelberg},\
  \bibinfo {year} {1999})\ pp.\ \bibinfo {pages} {473--480}\BibitemShut
  {NoStop}%
\bibitem [{\citenamefont {Parzen}(1962)}]{parzen_kde}%
  \BibitemOpen
  \bibfield  {author} {\bibinfo {author} {\bibfnamefont {E.}~\bibnamefont
  {Parzen}},\ }\bibfield  {title} {\bibinfo {title} {On estimation of a
  probability density function and mode},\ }\href
  {https://doi.org/10.1214/aoms/1177704472} {\bibfield  {journal} {\bibinfo
  {journal} {Ann. Math. Statist.}\ }\textbf {\bibinfo {volume} {33}},\ \bibinfo
  {pages} {1065} (\bibinfo {year} {1962})}\BibitemShut {NoStop}%
\bibitem [{\citenamefont {Murphy}(2013)}]{murphy2013machine}%
  \BibitemOpen
  \bibfield  {author} {\bibinfo {author} {\bibfnamefont {K.~P.}\ \bibnamefont
  {Murphy}},\ }\href@noop {} {\emph {\bibinfo {title} {Machine learning : a
  probabilistic perspective}}}\ (\bibinfo  {publisher} {MIT Press},\ \bibinfo
  {address} {Cambridge, Mass. [u.a.]},\ \bibinfo {year} {2013})\BibitemShut
  {NoStop}%
\bibitem [{\citenamefont {Nozari}\ \emph {et~al.}(2020)\citenamefont {Nozari},
  \citenamefont {Stiso}, \citenamefont {Caciagli}, \citenamefont {Cornblath},
  \citenamefont {He}, \citenamefont {Bertolero}, \citenamefont {Mahadevan},
  \citenamefont {Pappas},\ and\ \citenamefont {Bassett}}]{nozari2021is}%
  \BibitemOpen
  \bibfield  {author} {\bibinfo {author} {\bibfnamefont {E.}~\bibnamefont
  {Nozari}}, \bibinfo {author} {\bibfnamefont {J.}~\bibnamefont {Stiso}},
  \bibinfo {author} {\bibfnamefont {L.}~\bibnamefont {Caciagli}}, \bibinfo
  {author} {\bibfnamefont {E.~J.}\ \bibnamefont {Cornblath}}, \bibinfo {author}
  {\bibfnamefont {X.}~\bibnamefont {He}}, \bibinfo {author} {\bibfnamefont
  {M.~A.}\ \bibnamefont {Bertolero}}, \bibinfo {author} {\bibfnamefont {A.~S.}\
  \bibnamefont {Mahadevan}}, \bibinfo {author} {\bibfnamefont {G.~J.}\
  \bibnamefont {Pappas}},\ and\ \bibinfo {author} {\bibfnamefont {D.~S.}\
  \bibnamefont {Bassett}},\ }\bibfield  {title} {\bibinfo {title} {Is the brain
  macroscopically linear? {A} system identification of resting state
  dynamics},\ }\href@noop {} {\bibfield  {journal} {\bibinfo  {journal}
  {arXiv}\ ,\ \bibinfo {pages} {12351}} (\bibinfo {year} {2020})}\BibitemShut
  {NoStop}%
\bibitem [{\citenamefont {Tang}\ and\ \citenamefont
  {Bassett}(2018)}]{colloq_cdbn}%
  \BibitemOpen
  \bibfield  {author} {\bibinfo {author} {\bibfnamefont {E.}~\bibnamefont
  {Tang}}\ and\ \bibinfo {author} {\bibfnamefont {D.~S.}\ \bibnamefont
  {Bassett}},\ }\bibfield  {title} {\bibinfo {title} {Colloqium: Control of
  dynamics in brain networks},\ }\bibfield  {journal} {\bibinfo  {journal}
  {Reviews of modern physics}\ }\textbf {\bibinfo {volume} {90}},\ \href
  {https://doi.org/10.1103/RevModPhys.90.031003} {10.1103/RevModPhys.90.031003}
  (\bibinfo {year} {2018})\BibitemShut {NoStop}%
\bibitem [{\citenamefont {Karrer}\ \emph {et~al.}(2020)\citenamefont {Karrer},
  \citenamefont {Kim}, \citenamefont {Stiso}, \citenamefont {Kahn},
  \citenamefont {Pasqualetti}, \citenamefont {Habel},\ and\ \citenamefont
  {Bassett}}]{karrer2019}%
  \BibitemOpen
  \bibfield  {author} {\bibinfo {author} {\bibfnamefont {T.}~\bibnamefont
  {Karrer}}, \bibinfo {author} {\bibfnamefont {J.}~\bibnamefont {Kim}},
  \bibinfo {author} {\bibfnamefont {J.}~\bibnamefont {Stiso}}, \bibinfo
  {author} {\bibfnamefont {A.}~\bibnamefont {Kahn}}, \bibinfo {author}
  {\bibfnamefont {F.}~\bibnamefont {Pasqualetti}}, \bibinfo {author}
  {\bibfnamefont {U.}~\bibnamefont {Habel}},\ and\ \bibinfo {author}
  {\bibfnamefont {D.}~\bibnamefont {Bassett}},\ }\bibfield  {title} {\bibinfo
  {title} {A practical guide to methodological considerations in the
  controllability of structural brain networks},\ }\bibfield  {journal}
  {\bibinfo  {journal} {Journal of Neural Engineering}\ }\textbf {\bibinfo
  {volume} {17}},\ \href {https://doi.org/10.1088/1741-2552/ab6e8b}
  {10.1088/1741-2552/ab6e8b} (\bibinfo {year} {2020})\BibitemShut {NoStop}%
\bibitem [{\citenamefont {Laughlin}(2001)}]{laughlin2001energy}%
  \BibitemOpen
  \bibfield  {author} {\bibinfo {author} {\bibfnamefont {S.~B.}\ \bibnamefont
  {Laughlin}},\ }\bibfield  {title} {\bibinfo {title} {Energy as a constraint
  on the coding and processing of sensory information},\ }\href@noop {}
  {\bibfield  {journal} {\bibinfo  {journal} {Curr Opin Neurobiol}\ }\textbf
  {\bibinfo {volume} {11}},\ \bibinfo {pages} {475} (\bibinfo {year}
  {2001})}\BibitemShut {NoStop}%
\bibitem [{\citenamefont {{Marx}}\ \emph {et~al.}(2004)\citenamefont {{Marx}},
  \citenamefont {{Koenig}},\ and\ \citenamefont
  {{Georges}}}]{avg_controllability1}%
  \BibitemOpen
  \bibfield  {author} {\bibinfo {author} {\bibfnamefont {B.}~\bibnamefont
  {{Marx}}}, \bibinfo {author} {\bibfnamefont {D.}~\bibnamefont {{Koenig}}},\
  and\ \bibinfo {author} {\bibfnamefont {D.}~\bibnamefont {{Georges}}},\
  }\bibfield  {title} {\bibinfo {title} {Optimal sensor and actuator location
  for descriptor systems using generalized gramians and balanced
  realizations},\ }in\ \href {https://doi.org/10.23919/ACC.2004.1383878} {\emph
  {\bibinfo {booktitle} {Proceedings of the 2004 American Control
  Conference}}},\ Vol.~\bibinfo {volume} {3}\ (\bibinfo {year} {2004})\ pp.\
  \bibinfo {pages} {2729--2734 vol.3}\BibitemShut {NoStop}%
\bibitem [{\citenamefont {Barch}\ \emph
  {et~al.}(2013{\natexlab{b}})\citenamefont {Barch}, \citenamefont {Burgess},
  \citenamefont {Harms}, \citenamefont {Petersen}, \citenamefont {Schlaggar},
  \citenamefont {Corbetta}, \citenamefont {Glasser}, \citenamefont {Curtiss},
  \citenamefont {Dixit}, \citenamefont {Feldt}, \citenamefont {Nolan},
  \citenamefont {Bryant}, \citenamefont {Hartley}, \citenamefont {Footer},
  \citenamefont {Bjork}, \citenamefont {Poldrack}, \citenamefont {Smith},
  \citenamefont {Johansen-Berg}, \citenamefont {Snyder},\ and\ \citenamefont
  {{Van Essen}}}]{hcp_tfmri}%
  \BibitemOpen
  \bibfield  {author} {\bibinfo {author} {\bibfnamefont {D.~M.}\ \bibnamefont
  {Barch}}, \bibinfo {author} {\bibfnamefont {G.~C.}\ \bibnamefont {Burgess}},
  \bibinfo {author} {\bibfnamefont {M.~P.}\ \bibnamefont {Harms}}, \bibinfo
  {author} {\bibfnamefont {S.~E.}\ \bibnamefont {Petersen}}, \bibinfo {author}
  {\bibfnamefont {B.~L.}\ \bibnamefont {Schlaggar}}, \bibinfo {author}
  {\bibfnamefont {M.}~\bibnamefont {Corbetta}}, \bibinfo {author}
  {\bibfnamefont {M.~F.}\ \bibnamefont {Glasser}}, \bibinfo {author}
  {\bibfnamefont {S.}~\bibnamefont {Curtiss}}, \bibinfo {author} {\bibfnamefont
  {S.}~\bibnamefont {Dixit}}, \bibinfo {author} {\bibfnamefont
  {C.}~\bibnamefont {Feldt}}, \bibinfo {author} {\bibfnamefont
  {D.}~\bibnamefont {Nolan}}, \bibinfo {author} {\bibfnamefont
  {E.}~\bibnamefont {Bryant}}, \bibinfo {author} {\bibfnamefont
  {T.}~\bibnamefont {Hartley}}, \bibinfo {author} {\bibfnamefont
  {O.}~\bibnamefont {Footer}}, \bibinfo {author} {\bibfnamefont {J.~M.}\
  \bibnamefont {Bjork}}, \bibinfo {author} {\bibfnamefont {R.}~\bibnamefont
  {Poldrack}}, \bibinfo {author} {\bibfnamefont {S.}~\bibnamefont {Smith}},
  \bibinfo {author} {\bibfnamefont {H.}~\bibnamefont {Johansen-Berg}}, \bibinfo
  {author} {\bibfnamefont {A.~Z.}\ \bibnamefont {Snyder}},\ and\ \bibinfo
  {author} {\bibfnamefont {D.~C.}\ \bibnamefont {{Van Essen}}},\ }\bibfield
  {title} {\bibinfo {title} {Function in the human connectome: Task-fmri and
  individual differences in behavior},\ }\href
  {https://doi.org/https://doi.org/10.1016/j.neuroimage.2013.05.033} {\bibfield
   {journal} {\bibinfo  {journal} {NeuroImage}\ }\textbf {\bibinfo {volume}
  {80}},\ \bibinfo {pages} {169} (\bibinfo {year} {2013}{\natexlab{b}})},\
  \bibinfo {note} {mapping the Connectome}\BibitemShut {NoStop}%
\bibitem [{\citenamefont {Yarkoni}\ \emph {et~al.}(2011)\citenamefont
  {Yarkoni}, \citenamefont {Poldrack}, \citenamefont {Nichols}, \citenamefont
  {Van~Essen},\ and\ \citenamefont {Wager}}]{neurosynth}%
  \BibitemOpen
  \bibfield  {author} {\bibinfo {author} {\bibfnamefont {T.}~\bibnamefont
  {Yarkoni}}, \bibinfo {author} {\bibfnamefont {R.~A.}\ \bibnamefont
  {Poldrack}}, \bibinfo {author} {\bibfnamefont {T.~E.}\ \bibnamefont
  {Nichols}}, \bibinfo {author} {\bibfnamefont {D.~C.}\ \bibnamefont
  {Van~Essen}},\ and\ \bibinfo {author} {\bibfnamefont {T.~D.}\ \bibnamefont
  {Wager}},\ }\bibfield  {title} {\bibinfo {title} {Large-scale automated
  synthesis of human functional neuroimaging data},\ }\href@noop {} {\bibfield
  {journal} {\bibinfo  {journal} {Nature methods}\ }\textbf {\bibinfo {volume}
  {8}},\ \bibinfo {pages} {665} (\bibinfo {year} {2011})}\BibitemShut {NoStop}%
\bibitem [{\citenamefont {Bijsterbosch}\ \emph {et~al.}(2017)\citenamefont
  {Bijsterbosch}, \citenamefont {Harrison}, \citenamefont {Duff}, \citenamefont
  {Alfaro-Almagro}, \citenamefont {Woolrich},\ and\ \citenamefont
  {Smith}}]{fMRI_amp_fluct}%
  \BibitemOpen
  \bibfield  {author} {\bibinfo {author} {\bibfnamefont {J.}~\bibnamefont
  {Bijsterbosch}}, \bibinfo {author} {\bibfnamefont {S.}~\bibnamefont
  {Harrison}}, \bibinfo {author} {\bibfnamefont {E.}~\bibnamefont {Duff}},
  \bibinfo {author} {\bibfnamefont {F.}~\bibnamefont {Alfaro-Almagro}},
  \bibinfo {author} {\bibfnamefont {M.}~\bibnamefont {Woolrich}},\ and\
  \bibinfo {author} {\bibfnamefont {S.}~\bibnamefont {Smith}},\ }\bibfield
  {title} {\bibinfo {title} {Investigations into within- and between-subject
  resting-state amplitude variations},\ }\href
  {https://doi.org/https://doi.org/10.1016/j.neuroimage.2017.07.014} {\bibfield
   {journal} {\bibinfo  {journal} {NeuroImage}\ }\textbf {\bibinfo {volume}
  {159}},\ \bibinfo {pages} {57} (\bibinfo {year} {2017})}\BibitemShut
  {NoStop}%
\bibitem [{\citenamefont {Whitfield-Gabrieli}\ and\ \citenamefont
  {Ford}(2012)}]{whitfield2012default}%
  \BibitemOpen
  \bibfield  {author} {\bibinfo {author} {\bibfnamefont {S.}~\bibnamefont
  {Whitfield-Gabrieli}}\ and\ \bibinfo {author} {\bibfnamefont {J.~M.}\
  \bibnamefont {Ford}},\ }\bibfield  {title} {\bibinfo {title} {Default mode
  network activity and connectivity in psychopathology},\ }\href@noop {}
  {\bibfield  {journal} {\bibinfo  {journal} {Annual review of clinical
  psychology}\ }\textbf {\bibinfo {volume} {8}},\ \bibinfo {pages} {49}
  (\bibinfo {year} {2012})}\BibitemShut {NoStop}%
\bibitem [{\citenamefont {Ito}\ \emph {et~al.}(2020)\citenamefont {Ito},
  \citenamefont {Brincat}, \citenamefont {Siegel}, \citenamefont {Mill},
  \citenamefont {He}, \citenamefont {Miller}, \citenamefont {Rotstein},\ and\
  \citenamefont {Cole}}]{ito2020task}%
  \BibitemOpen
  \bibfield  {author} {\bibinfo {author} {\bibfnamefont {T.}~\bibnamefont
  {Ito}}, \bibinfo {author} {\bibfnamefont {S.~L.}\ \bibnamefont {Brincat}},
  \bibinfo {author} {\bibfnamefont {M.}~\bibnamefont {Siegel}}, \bibinfo
  {author} {\bibfnamefont {R.~D.}\ \bibnamefont {Mill}}, \bibinfo {author}
  {\bibfnamefont {B.~J.}\ \bibnamefont {He}}, \bibinfo {author} {\bibfnamefont
  {E.~K.}\ \bibnamefont {Miller}}, \bibinfo {author} {\bibfnamefont {H.~G.}\
  \bibnamefont {Rotstein}},\ and\ \bibinfo {author} {\bibfnamefont {M.~W.}\
  \bibnamefont {Cole}},\ }\bibfield  {title} {\bibinfo {title} {Task-evoked
  activity quenches neural correlations and variability across cortical
  areas},\ }\href@noop {} {\bibfield  {journal} {\bibinfo  {journal} {PLoS
  computational biology}\ }\textbf {\bibinfo {volume} {16}},\ \bibinfo {pages}
  {e1007983} (\bibinfo {year} {2020})}\BibitemShut {NoStop}%
\bibitem [{\citenamefont {Amico}\ \emph {et~al.}(2020)\citenamefont {Amico},
  \citenamefont {Abbas}, \citenamefont {Duong-Tran}, \citenamefont {Tipnis},
  \citenamefont {Rajapandian}, \citenamefont {Chumin}, \citenamefont
  {Ventresca}, \citenamefont {Harezlak},\ and\ \citenamefont
  {Goñi}}]{amico2020information}%
  \BibitemOpen
  \bibfield  {author} {\bibinfo {author} {\bibfnamefont {E.}~\bibnamefont
  {Amico}}, \bibinfo {author} {\bibfnamefont {K.}~\bibnamefont {Abbas}},
  \bibinfo {author} {\bibfnamefont {D.~A.}\ \bibnamefont {Duong-Tran}},
  \bibinfo {author} {\bibfnamefont {U.}~\bibnamefont {Tipnis}}, \bibinfo
  {author} {\bibfnamefont {M.}~\bibnamefont {Rajapandian}}, \bibinfo {author}
  {\bibfnamefont {E.}~\bibnamefont {Chumin}}, \bibinfo {author} {\bibfnamefont
  {M.}~\bibnamefont {Ventresca}}, \bibinfo {author} {\bibfnamefont
  {J.}~\bibnamefont {Harezlak}},\ and\ \bibinfo {author} {\bibfnamefont
  {J.}~\bibnamefont {Goñi}},\ }\href@noop {} {\bibinfo {title} {Towards an
  information theoretical description of communication in brain networks}}
  (\bibinfo {year} {2020}),\ \Eprint {https://arxiv.org/abs/1911.02601}
  {1911.02601} \BibitemShut {NoStop}%
\bibitem [{\citenamefont {Ponce-Alvarez}\ \emph {et~al.}(2015)\citenamefont
  {Ponce-Alvarez}, \citenamefont {He}, \citenamefont {Hagmann},\ and\
  \citenamefont {Deco}}]{ponce2015task}%
  \BibitemOpen
  \bibfield  {author} {\bibinfo {author} {\bibfnamefont {A.}~\bibnamefont
  {Ponce-Alvarez}}, \bibinfo {author} {\bibfnamefont {B.~J.}\ \bibnamefont
  {He}}, \bibinfo {author} {\bibfnamefont {P.}~\bibnamefont {Hagmann}},\ and\
  \bibinfo {author} {\bibfnamefont {G.}~\bibnamefont {Deco}},\ }\bibfield
  {title} {\bibinfo {title} {Task-driven activity reduces the cortical activity
  space of the brain: experiment and whole-brain modeling},\ }\href@noop {}
  {\bibfield  {journal} {\bibinfo  {journal} {PLoS Comput Biol}\ }\textbf
  {\bibinfo {volume} {11}},\ \bibinfo {pages} {e1004445} (\bibinfo {year}
  {2015})}\BibitemShut {NoStop}%
\bibitem [{\citenamefont {Timme}\ and\ \citenamefont
  {Lapish}(2018)}]{itn_tutorial}%
  \BibitemOpen
  \bibfield  {author} {\bibinfo {author} {\bibfnamefont {N.~M.}\ \bibnamefont
  {Timme}}\ and\ \bibinfo {author} {\bibfnamefont {C.}~\bibnamefont {Lapish}},\
  }\bibfield  {title} {\bibinfo {title} {A tutorial for information theory in
  neuroscience},\ }\bibfield  {journal} {\bibinfo  {journal} {eNeuro}\ }\textbf
  {\bibinfo {volume} {5}},\ \href {https://doi.org/10.1523/ENEURO.0052-18.2018}
  {10.1523/ENEURO.0052-18.2018} (\bibinfo {year} {2018})\BibitemShut {NoStop}%
\bibitem [{\citenamefont {Reinagel}(2000)}]{it_reinagel}%
  \BibitemOpen
  \bibfield  {author} {\bibinfo {author} {\bibfnamefont {P.}~\bibnamefont
  {Reinagel}},\ }\bibfield  {title} {\bibinfo {title} {Information theory in
  the brain},\ }\href@noop {} {\bibfield  {journal} {\bibinfo  {journal}
  {Current Biology}\ }\textbf {\bibinfo {volume} {10}},\ \bibinfo {pages}
  {R542} (\bibinfo {year} {2000})}\BibitemShut {NoStop}%
\bibitem [{\citenamefont {Rabinovich}\ \emph {et~al.}(2012)\citenamefont
  {Rabinovich}, \citenamefont {Friston},\ and\ \citenamefont
  {Varona}}]{rabinovich2012principles}%
  \BibitemOpen
  \bibfield  {author} {\bibinfo {author} {\bibfnamefont {M.~I.}\ \bibnamefont
  {Rabinovich}}, \bibinfo {author} {\bibfnamefont {K.~J.}\ \bibnamefont
  {Friston}},\ and\ \bibinfo {author} {\bibfnamefont {P.}~\bibnamefont
  {Varona}},\ }\href@noop {} {\emph {\bibinfo {title} {Principles of brain
  dynamics: global state interactions}}}\ (\bibinfo  {publisher} {MIT Press},\
  \bibinfo {year} {2012})\BibitemShut {NoStop}%
\bibitem [{\citenamefont {Fan}(2014)}]{itacc_fan}%
  \BibitemOpen
  \bibfield  {author} {\bibinfo {author} {\bibfnamefont {J.}~\bibnamefont
  {Fan}},\ }\bibfield  {title} {\bibinfo {title} {An information theory account
  of cognitive control},\ }\href {https://doi.org/10.3389/fnhum.2014.00680}
  {\bibfield  {journal} {\bibinfo  {journal} {Frontiers in Human Neuroscience}\
  }\textbf {\bibinfo {volume} {8}},\ \bibinfo {pages} {680} (\bibinfo {year}
  {2014})}\BibitemShut {NoStop}%
\bibitem [{\citenamefont {Nolte}\ \emph {et~al.}(2008)\citenamefont {Nolte},
  \citenamefont {Ziehe}, \citenamefont {Nikulin}, \citenamefont {Schl\"ogl},
  \citenamefont {Kr\"amer}, \citenamefont {Brismar},\ and\ \citenamefont
  {M\"uller}}]{est_flow_info}%
  \BibitemOpen
  \bibfield  {author} {\bibinfo {author} {\bibfnamefont {G.}~\bibnamefont
  {Nolte}}, \bibinfo {author} {\bibfnamefont {A.}~\bibnamefont {Ziehe}},
  \bibinfo {author} {\bibfnamefont {V.~V.}\ \bibnamefont {Nikulin}}, \bibinfo
  {author} {\bibfnamefont {A.}~\bibnamefont {Schl\"ogl}}, \bibinfo {author}
  {\bibfnamefont {N.}~\bibnamefont {Kr\"amer}}, \bibinfo {author}
  {\bibfnamefont {T.}~\bibnamefont {Brismar}},\ and\ \bibinfo {author}
  {\bibfnamefont {K.-R.}\ \bibnamefont {M\"uller}},\ }\bibfield  {title}
  {\bibinfo {title} {Robustly estimating the flow direction of information in
  complex physical systems},\ }\href
  {https://doi.org/10.1103/PhysRevLett.100.234101} {\bibfield  {journal}
  {\bibinfo  {journal} {Phys. Rev. Lett.}\ }\textbf {\bibinfo {volume} {100}},\
  \bibinfo {pages} {234101} (\bibinfo {year} {2008})}\BibitemShut {NoStop}%
\bibitem [{\citenamefont {Keshmiri}(2020)}]{e22090917}%
  \BibitemOpen
  \bibfield  {author} {\bibinfo {author} {\bibfnamefont {S.}~\bibnamefont
  {Keshmiri}},\ }\bibfield  {title} {\bibinfo {title} {Entropy and the brain:
  An overview},\ }\bibfield  {journal} {\bibinfo  {journal} {Entropy}\ }\textbf
  {\bibinfo {volume} {22}},\ \href {https://doi.org/10.3390/e22090917}
  {10.3390/e22090917} (\bibinfo {year} {2020})\BibitemShut {NoStop}%
\bibitem [{\citenamefont {Saxe}\ \emph {et~al.}(2018)\citenamefont {Saxe},
  \citenamefont {Calderone},\ and\ \citenamefont {Morales}}]{entr_intel18}%
  \BibitemOpen
  \bibfield  {author} {\bibinfo {author} {\bibfnamefont {G.~N.}\ \bibnamefont
  {Saxe}}, \bibinfo {author} {\bibfnamefont {D.}~\bibnamefont {Calderone}},\
  and\ \bibinfo {author} {\bibfnamefont {L.~J.}\ \bibnamefont {Morales}},\
  }\bibfield  {title} {\bibinfo {title} {Brain entropy and human intelligence:
  A resting-state fmri study},\ }\href
  {https://doi.org/10.1371/journal.pone.0191582} {\bibfield  {journal}
  {\bibinfo  {journal} {PLOS ONE}\ }\textbf {\bibinfo {volume} {13}},\ \bibinfo
  {pages} {1} (\bibinfo {year} {2018})}\BibitemShut {NoStop}%
\bibitem [{\citenamefont {Wang}\ \emph {et~al.}(2014)\citenamefont {Wang},
  \citenamefont {Li}, \citenamefont {Childress},\ and\ \citenamefont
  {Detre}}]{bentr_fmri14}%
  \BibitemOpen
  \bibfield  {author} {\bibinfo {author} {\bibfnamefont {Z.}~\bibnamefont
  {Wang}}, \bibinfo {author} {\bibfnamefont {Y.}~\bibnamefont {Li}}, \bibinfo
  {author} {\bibfnamefont {A.~R.}\ \bibnamefont {Childress}},\ and\ \bibinfo
  {author} {\bibfnamefont {J.~A.}\ \bibnamefont {Detre}},\ }\bibfield  {title}
  {\bibinfo {title} {Brain entropy mapping using fmri},\ }\href
  {https://doi.org/10.1371/journal.pone.0089948} {\bibfield  {journal}
  {\bibinfo  {journal} {PLOS ONE}\ }\textbf {\bibinfo {volume} {9}},\ \bibinfo
  {pages} {1} (\bibinfo {year} {2014})}\BibitemShut {NoStop}%
\bibitem [{\citenamefont {Liu}\ \emph {et~al.}(2020)\citenamefont {Liu},
  \citenamefont {Liu}, \citenamefont {Hildebrandt},\ and\ \citenamefont
  {Zhou}}]{icep_liu}%
  \BibitemOpen
  \bibfield  {author} {\bibinfo {author} {\bibfnamefont {M.}~\bibnamefont
  {Liu}}, \bibinfo {author} {\bibfnamefont {X.}~\bibnamefont {Liu}}, \bibinfo
  {author} {\bibfnamefont {A.}~\bibnamefont {Hildebrandt}},\ and\ \bibinfo
  {author} {\bibfnamefont {C.}~\bibnamefont {Zhou}},\ }\bibfield  {title}
  {\bibinfo {title} {{Individual Cortical Entropy Profile: Test–Retest
  Reliability, Predictive Power for Cognitive Ability, and Neuroanatomical
  Foundation}},\ }\bibfield  {journal} {\bibinfo  {journal} {Cerebral Cortex
  Communications}\ }\textbf {\bibinfo {volume} {1}},\ \href
  {https://doi.org/10.1093/texcom/tgaa015} {10.1093/texcom/tgaa015} (\bibinfo
  {year} {2020})\BibitemShut {NoStop}%
\bibitem [{\citenamefont {Chai}\ \emph {et~al.}(2009)\citenamefont {Chai},
  \citenamefont {Walther}, \citenamefont {Beck},\ and\ \citenamefont
  {Fei-fei}}]{mi_fmri_knn}%
  \BibitemOpen
  \bibfield  {author} {\bibinfo {author} {\bibfnamefont {B.}~\bibnamefont
  {Chai}}, \bibinfo {author} {\bibfnamefont {D.}~\bibnamefont {Walther}},
  \bibinfo {author} {\bibfnamefont {D.}~\bibnamefont {Beck}},\ and\ \bibinfo
  {author} {\bibfnamefont {L.}~\bibnamefont {Fei-fei}},\ }\bibfield  {title}
  {\bibinfo {title} {Exploring functional connectivities of the human brain
  using multivariate information analysis},\ }in\ \href@noop {} {\emph
  {\bibinfo {booktitle} {Advances in Neural Information Processing Systems}}},\
  Vol.~\bibinfo {volume} {22},\ \bibinfo {editor} {edited by\ \bibinfo {editor}
  {\bibfnamefont {Y.}~\bibnamefont {Bengio}}, \bibinfo {editor} {\bibfnamefont
  {D.}~\bibnamefont {Schuurmans}}, \bibinfo {editor} {\bibfnamefont
  {J.}~\bibnamefont {Lafferty}}, \bibinfo {editor} {\bibfnamefont
  {C.}~\bibnamefont {Williams}},\ and\ \bibinfo {editor} {\bibfnamefont
  {A.}~\bibnamefont {Culotta}}}\ (\bibinfo  {publisher} {Curran Associates,
  Inc.},\ \bibinfo {year} {2009})\ pp.\ \bibinfo {pages} {270--278}\BibitemShut
  {NoStop}%
\bibitem [{\citenamefont {Tang}\ \emph {et~al.}(2017)\citenamefont {Tang},
  \citenamefont {Giusti}, \citenamefont {Baum}, \citenamefont {Gu},
  \citenamefont {Pollock}, \citenamefont {Kahn}, \citenamefont {Roalf},
  \citenamefont {Moore}, \citenamefont {Ruparel}, \citenamefont {Gur} \emph
  {et~al.}}]{tang2017developmental}%
  \BibitemOpen
  \bibfield  {author} {\bibinfo {author} {\bibfnamefont {E.}~\bibnamefont
  {Tang}}, \bibinfo {author} {\bibfnamefont {C.}~\bibnamefont {Giusti}},
  \bibinfo {author} {\bibfnamefont {G.~L.}\ \bibnamefont {Baum}}, \bibinfo
  {author} {\bibfnamefont {S.}~\bibnamefont {Gu}}, \bibinfo {author}
  {\bibfnamefont {E.}~\bibnamefont {Pollock}}, \bibinfo {author} {\bibfnamefont
  {A.~E.}\ \bibnamefont {Kahn}}, \bibinfo {author} {\bibfnamefont {D.~R.}\
  \bibnamefont {Roalf}}, \bibinfo {author} {\bibfnamefont {T.~M.}\ \bibnamefont
  {Moore}}, \bibinfo {author} {\bibfnamefont {K.}~\bibnamefont {Ruparel}},
  \bibinfo {author} {\bibfnamefont {R.~C.}\ \bibnamefont {Gur}}, \emph
  {et~al.},\ }\bibfield  {title} {\bibinfo {title} {Developmental increases in
  white matter network controllability support a growing diversity of brain
  dynamics},\ }\href@noop {} {\bibfield  {journal} {\bibinfo  {journal} {Nature
  communications}\ }\textbf {\bibinfo {volume} {8}},\ \bibinfo {pages} {1}
  (\bibinfo {year} {2017})}\BibitemShut {NoStop}%
\bibitem [{\citenamefont {Braun}\ \emph {et~al.}(2019)\citenamefont {Braun},
  \citenamefont {Harneit}, \citenamefont {Pergola}, \citenamefont {Menara},
  \citenamefont {Schaefer}, \citenamefont {Betzel}, \citenamefont {Zang},
  \citenamefont {Schweiger}, \citenamefont {Schwarz}, \citenamefont {Chen}
  \emph {et~al.}}]{braun2019brain}%
  \BibitemOpen
  \bibfield  {author} {\bibinfo {author} {\bibfnamefont {U.}~\bibnamefont
  {Braun}}, \bibinfo {author} {\bibfnamefont {A.}~\bibnamefont {Harneit}},
  \bibinfo {author} {\bibfnamefont {G.}~\bibnamefont {Pergola}}, \bibinfo
  {author} {\bibfnamefont {T.}~\bibnamefont {Menara}}, \bibinfo {author}
  {\bibfnamefont {A.}~\bibnamefont {Schaefer}}, \bibinfo {author}
  {\bibfnamefont {R.~F.}\ \bibnamefont {Betzel}}, \bibinfo {author}
  {\bibfnamefont {Z.}~\bibnamefont {Zang}}, \bibinfo {author} {\bibfnamefont
  {J.~I.}\ \bibnamefont {Schweiger}}, \bibinfo {author} {\bibfnamefont
  {K.}~\bibnamefont {Schwarz}}, \bibinfo {author} {\bibfnamefont
  {J.}~\bibnamefont {Chen}}, \emph {et~al.},\ }\bibfield  {title} {\bibinfo
  {title} {Brain state stability during working memory is explained by network
  control theory, modulated by dopamine d1/d2 receptor function, and diminished
  in schizophrenia},\ }\href@noop {} {\bibfield  {journal} {\bibinfo  {journal}
  {arXiv preprint arXiv:1906.09290}\ } (\bibinfo {year} {2019})}\BibitemShut
  {NoStop}%
\bibitem [{\citenamefont {Bullmore}\ and\ \citenamefont
  {Sporns}(2012)}]{bullmore2012economy}%
  \BibitemOpen
  \bibfield  {author} {\bibinfo {author} {\bibfnamefont {E.}~\bibnamefont
  {Bullmore}}\ and\ \bibinfo {author} {\bibfnamefont {O.}~\bibnamefont
  {Sporns}},\ }\bibfield  {title} {\bibinfo {title} {The economy of brain
  network organization},\ }\href@noop {} {\bibfield  {journal} {\bibinfo
  {journal} {Nature Reviews Neuroscience}\ }\textbf {\bibinfo {volume} {13}},\
  \bibinfo {pages} {336} (\bibinfo {year} {2012})}\BibitemShut {NoStop}%
\bibitem [{\citenamefont {MacKay}\ and\ \citenamefont
  {Mac~Kay}(2003)}]{mackay2003information}%
  \BibitemOpen
  \bibfield  {author} {\bibinfo {author} {\bibfnamefont {D.~J.}\ \bibnamefont
  {MacKay}}\ and\ \bibinfo {author} {\bibfnamefont {D.~J.}\ \bibnamefont
  {Mac~Kay}},\ }\href@noop {} {\emph {\bibinfo {title} {Information theory,
  inference and learning algorithms}}}\ (\bibinfo  {publisher} {Cambridge
  university press},\ \bibinfo {year} {2003})\BibitemShut {NoStop}%
\bibitem [{\citenamefont {Bain}(1950)}]{zipf2016human}%
  \BibitemOpen
  \bibfield  {author} {\bibinfo {author} {\bibfnamefont {R.}~\bibnamefont
  {Bain}},\ }\bibfield  {title} {\bibinfo {title} {{Human Behavior and the
  Principle of Least Effort: An Introduction to Human Ecology. By George
  Kingsley Zipf. Cambridge, Mass.: Addison-Wesley Press, Inc., 1949. 573 pp.
  6.50}},\ }\href {https://doi.org/10.2307/2572028} {\bibfield  {journal}
  {\bibinfo  {journal} {Social Forces}\ }\textbf {\bibinfo {volume} {28}},\
  \bibinfo {pages} {340} (\bibinfo {year} {1950})}\BibitemShut {NoStop}%
\bibitem [{\citenamefont {Zhou}\ \emph
  {et~al.}(2020{\natexlab{a}})\citenamefont {Zhou}, \citenamefont {Lynn},
  \citenamefont {Cui}, \citenamefont {Ciric}, \citenamefont {Baum},
  \citenamefont {Moore}, \citenamefont {Roalf}, \citenamefont {Detre},
  \citenamefont {Gur}, \citenamefont {Gur}, \citenamefont {Satterthwaite},\
  and\ \citenamefont {Bassett}}]{zhou2020efficient}%
  \BibitemOpen
  \bibfield  {author} {\bibinfo {author} {\bibfnamefont {D.}~\bibnamefont
  {Zhou}}, \bibinfo {author} {\bibfnamefont {C.~W.}\ \bibnamefont {Lynn}},
  \bibinfo {author} {\bibfnamefont {Z.}~\bibnamefont {Cui}}, \bibinfo {author}
  {\bibfnamefont {R.}~\bibnamefont {Ciric}}, \bibinfo {author} {\bibfnamefont
  {G.~L.}\ \bibnamefont {Baum}}, \bibinfo {author} {\bibfnamefont {T.~M.}\
  \bibnamefont {Moore}}, \bibinfo {author} {\bibfnamefont {D.~R.}\ \bibnamefont
  {Roalf}}, \bibinfo {author} {\bibfnamefont {J.~A.}\ \bibnamefont {Detre}},
  \bibinfo {author} {\bibfnamefont {R.~C.}\ \bibnamefont {Gur}}, \bibinfo
  {author} {\bibfnamefont {R.~E.}\ \bibnamefont {Gur}}, \bibinfo {author}
  {\bibfnamefont {T.~D.}\ \bibnamefont {Satterthwaite}},\ and\ \bibinfo
  {author} {\bibfnamefont {D.~S.}\ \bibnamefont {Bassett}},\ }\bibfield
  {title} {\bibinfo {title} {Efficient coding in the economics of human brain
  connectomics},\ }\href@noop {} {\bibfield  {journal} {\bibinfo  {journal}
  {bioRxiv}\ } (\bibinfo {year} {2020}{\natexlab{a}})},\ \Eprint
  {https://arxiv.org/abs/2001.05078} {2001.05078} \BibitemShut {NoStop}%
\bibitem [{\citenamefont {Ercsey-Ravasz}\ \emph {et~al.}(2013)\citenamefont
  {Ercsey-Ravasz}, \citenamefont {Markov}, \citenamefont {Lamy}, \citenamefont
  {Van~Essen}, \citenamefont {Knoblauch}, \citenamefont {Toroczkai},\ and\
  \citenamefont {Kennedy}}]{ercsey2013predictive}%
  \BibitemOpen
  \bibfield  {author} {\bibinfo {author} {\bibfnamefont {M.}~\bibnamefont
  {Ercsey-Ravasz}}, \bibinfo {author} {\bibfnamefont {N.~T.}\ \bibnamefont
  {Markov}}, \bibinfo {author} {\bibfnamefont {C.}~\bibnamefont {Lamy}},
  \bibinfo {author} {\bibfnamefont {D.~C.}\ \bibnamefont {Van~Essen}}, \bibinfo
  {author} {\bibfnamefont {K.}~\bibnamefont {Knoblauch}}, \bibinfo {author}
  {\bibfnamefont {Z.}~\bibnamefont {Toroczkai}},\ and\ \bibinfo {author}
  {\bibfnamefont {H.}~\bibnamefont {Kennedy}},\ }\bibfield  {title} {\bibinfo
  {title} {A predictive network model of cerebral cortical connectivity based
  on a distance rule},\ }\href@noop {} {\bibfield  {journal} {\bibinfo
  {journal} {Neuron}\ }\textbf {\bibinfo {volume} {80}},\ \bibinfo {pages}
  {184} (\bibinfo {year} {2013})}\BibitemShut {NoStop}%
\bibitem [{\citenamefont {Stiso}\ and\ \citenamefont
  {Bassett}(2018)}]{stiso2018spatial}%
  \BibitemOpen
  \bibfield  {author} {\bibinfo {author} {\bibfnamefont {J.}~\bibnamefont
  {Stiso}}\ and\ \bibinfo {author} {\bibfnamefont {D.~S.}\ \bibnamefont
  {Bassett}},\ }\bibfield  {title} {\bibinfo {title} {Spatial embedding imposes
  constraints on neuronal network architectures},\ }\href@noop {} {\bibfield
  {journal} {\bibinfo  {journal} {Trends in cognitive sciences}\ }\textbf
  {\bibinfo {volume} {22}},\ \bibinfo {pages} {1127} (\bibinfo {year}
  {2018})}\BibitemShut {NoStop}%
\bibitem [{\citenamefont {Kim}\ \emph {et~al.}(2018)\citenamefont {Kim},
  \citenamefont {Soffer}, \citenamefont {Kahn}, \citenamefont {Vettel},
  \citenamefont {Pasqualetti},\ and\ \citenamefont {Bassett}}]{kim2018role}%
  \BibitemOpen
  \bibfield  {author} {\bibinfo {author} {\bibfnamefont {J.~Z.}\ \bibnamefont
  {Kim}}, \bibinfo {author} {\bibfnamefont {J.~M.}\ \bibnamefont {Soffer}},
  \bibinfo {author} {\bibfnamefont {A.~E.}\ \bibnamefont {Kahn}}, \bibinfo
  {author} {\bibfnamefont {J.~M.}\ \bibnamefont {Vettel}}, \bibinfo {author}
  {\bibfnamefont {F.}~\bibnamefont {Pasqualetti}},\ and\ \bibinfo {author}
  {\bibfnamefont {D.~S.}\ \bibnamefont {Bassett}},\ }\bibfield  {title}
  {\bibinfo {title} {Role of graph architecture in controlling dynamical
  networks with applications to neural systems},\ }\href@noop {} {\bibfield
  {journal} {\bibinfo  {journal} {Nature physics}\ }\textbf {\bibinfo {volume}
  {14}},\ \bibinfo {pages} {91} (\bibinfo {year} {2018})}\BibitemShut {NoStop}%
\bibitem [{\citenamefont {He}\ \emph {et~al.}(2021)\citenamefont {He},
  \citenamefont {Caciagli}, \citenamefont {Parkes}, \citenamefont {Stiso},
  \citenamefont {Karrer}, \citenamefont {Kim}, \citenamefont {Lu},
  \citenamefont {Menara}, \citenamefont {Pasqualetti}, \citenamefont
  {Sperling}, \citenamefont {Tracy},\ and\ \citenamefont
  {Bassett}}]{control_metabolism}%
  \BibitemOpen
  \bibfield  {author} {\bibinfo {author} {\bibfnamefont {X.}~\bibnamefont
  {He}}, \bibinfo {author} {\bibfnamefont {L.}~\bibnamefont {Caciagli}},
  \bibinfo {author} {\bibfnamefont {L.}~\bibnamefont {Parkes}}, \bibinfo
  {author} {\bibfnamefont {J.}~\bibnamefont {Stiso}}, \bibinfo {author}
  {\bibfnamefont {T.~M.}\ \bibnamefont {Karrer}}, \bibinfo {author}
  {\bibfnamefont {J.~Z.}\ \bibnamefont {Kim}}, \bibinfo {author} {\bibfnamefont
  {Z.}~\bibnamefont {Lu}}, \bibinfo {author} {\bibfnamefont {T.}~\bibnamefont
  {Menara}}, \bibinfo {author} {\bibfnamefont {F.}~\bibnamefont {Pasqualetti}},
  \bibinfo {author} {\bibfnamefont {M.~R.}\ \bibnamefont {Sperling}}, \bibinfo
  {author} {\bibfnamefont {J.~I.}\ \bibnamefont {Tracy}},\ and\ \bibinfo
  {author} {\bibfnamefont {D.~S.}\ \bibnamefont {Bassett}},\ }\bibfield
  {title} {\bibinfo {title} {Pathological and metabolic underpinnings of
  energetic inefficiency in temporal lobe epilepsy},\ }\bibfield  {journal}
  {\bibinfo  {journal} {bioRxiv}\ }\href
  {https://doi.org/10.1101/2021.09.23.461495} {10.1101/2021.09.23.461495}
  (\bibinfo {year} {2021})\BibitemShut {NoStop}%
\bibitem [{\citenamefont {Su{\'a}rez}\ \emph {et~al.}(2020)\citenamefont
  {Su{\'a}rez}, \citenamefont {Markello}, \citenamefont {Betzel},\ and\
  \citenamefont {Misic}}]{suarez2020linking}%
  \BibitemOpen
  \bibfield  {author} {\bibinfo {author} {\bibfnamefont {L.~E.}\ \bibnamefont
  {Su{\'a}rez}}, \bibinfo {author} {\bibfnamefont {R.~D.}\ \bibnamefont
  {Markello}}, \bibinfo {author} {\bibfnamefont {R.~F.}\ \bibnamefont
  {Betzel}},\ and\ \bibinfo {author} {\bibfnamefont {B.}~\bibnamefont
  {Misic}},\ }\bibfield  {title} {\bibinfo {title} {Linking structure and
  function in macroscale brain networks},\ }\href@noop {} {\bibfield  {journal}
  {\bibinfo  {journal} {Trends in Cognitive Sciences}\ }\textbf {\bibinfo
  {volume} {24}},\ \bibinfo {pages} {302} (\bibinfo {year} {2020})}\BibitemShut
  {NoStop}%
\bibitem [{\citenamefont {He}(2013)}]{he2013spontaneous}%
  \BibitemOpen
  \bibfield  {author} {\bibinfo {author} {\bibfnamefont {B.~J.}\ \bibnamefont
  {He}},\ }\bibfield  {title} {\bibinfo {title} {Spontaneous and task-evoked
  brain activity negatively interact},\ }\href@noop {} {\bibfield  {journal}
  {\bibinfo  {journal} {Journal of Neuroscience}\ }\textbf {\bibinfo {volume}
  {33}},\ \bibinfo {pages} {4672} (\bibinfo {year} {2013})}\BibitemShut
  {NoStop}%
\bibitem [{\citenamefont {Mazzucato}\ \emph {et~al.}(2016)\citenamefont
  {Mazzucato}, \citenamefont {Fontanini},\ and\ \citenamefont
  {La~Camera}}]{mazzucato2016stimuli}%
  \BibitemOpen
  \bibfield  {author} {\bibinfo {author} {\bibfnamefont {L.}~\bibnamefont
  {Mazzucato}}, \bibinfo {author} {\bibfnamefont {A.}~\bibnamefont
  {Fontanini}},\ and\ \bibinfo {author} {\bibfnamefont {G.}~\bibnamefont
  {La~Camera}},\ }\bibfield  {title} {\bibinfo {title} {Stimuli reduce the
  dimensionality of cortical activity},\ }\href@noop {} {\bibfield  {journal}
  {\bibinfo  {journal} {Frontiers in systems neuroscience}\ }\textbf {\bibinfo
  {volume} {10}},\ \bibinfo {pages} {11} (\bibinfo {year} {2016})}\BibitemShut
  {NoStop}%
\bibitem [{\citenamefont {Baker}\ \emph {et~al.}(2019)\citenamefont {Baker},
  \citenamefont {Dillon}, \citenamefont {Patrick}, \citenamefont {Roffman},
  \citenamefont {Brady}, \citenamefont {Pizzagalli}, \citenamefont
  {{\"O}ng{\"u}r},\ and\ \citenamefont {Holmes}}]{baker52}%
  \BibitemOpen
  \bibfield  {author} {\bibinfo {author} {\bibfnamefont {J.~T.}\ \bibnamefont
  {Baker}}, \bibinfo {author} {\bibfnamefont {D.~G.}\ \bibnamefont {Dillon}},
  \bibinfo {author} {\bibfnamefont {L.~M.}\ \bibnamefont {Patrick}}, \bibinfo
  {author} {\bibfnamefont {J.~L.}\ \bibnamefont {Roffman}}, \bibinfo {author}
  {\bibfnamefont {R.~O.}\ \bibnamefont {Brady}}, \bibinfo {author}
  {\bibfnamefont {D.~A.}\ \bibnamefont {Pizzagalli}}, \bibinfo {author}
  {\bibfnamefont {D.}~\bibnamefont {{\"O}ng{\"u}r}},\ and\ \bibinfo {author}
  {\bibfnamefont {A.~J.}\ \bibnamefont {Holmes}},\ }\bibfield  {title}
  {\bibinfo {title} {Functional connectomics of affective and psychotic
  pathology},\ }\href {https://doi.org/10.1073/pnas.1820780116} {\bibfield
  {journal} {\bibinfo  {journal} {Proceedings of the National Academy of
  Sciences}\ }\textbf {\bibinfo {volume} {116}},\ \bibinfo {pages} {9050}
  (\bibinfo {year} {2019})}\BibitemShut {NoStop}%
\bibitem [{\citenamefont {Smith}\ \emph {et~al.}(2015)\citenamefont {Smith},
  \citenamefont {Nichols}, \citenamefont {Vidaurre}, \citenamefont {Winkler},
  \citenamefont {Behrens}, \citenamefont {Glasser}, \citenamefont {Ugurbil},
  \citenamefont {Barch}, \citenamefont {Van~Essen},\ and\ \citenamefont
  {Miller}}]{smith53}%
  \BibitemOpen
  \bibfield  {author} {\bibinfo {author} {\bibfnamefont {S.}~\bibnamefont
  {Smith}}, \bibinfo {author} {\bibfnamefont {T.}~\bibnamefont {Nichols}},
  \bibinfo {author} {\bibfnamefont {D.}~\bibnamefont {Vidaurre}}, \bibinfo
  {author} {\bibfnamefont {A.}~\bibnamefont {Winkler}}, \bibinfo {author}
  {\bibfnamefont {T.}~\bibnamefont {Behrens}}, \bibinfo {author} {\bibfnamefont
  {M.}~\bibnamefont {Glasser}}, \bibinfo {author} {\bibfnamefont
  {K.}~\bibnamefont {Ugurbil}}, \bibinfo {author} {\bibfnamefont
  {D.}~\bibnamefont {Barch}}, \bibinfo {author} {\bibfnamefont
  {D.}~\bibnamefont {Van~Essen}},\ and\ \bibinfo {author} {\bibfnamefont
  {K.}~\bibnamefont {Miller}},\ }\bibfield  {title} {\bibinfo {title} {A
  positive-negative mode of population covariation links brain connectivity,
  demographics and behavior},\ }\bibfield  {journal} {\bibinfo  {journal}
  {Nature neuroscience}\ }\textbf {\bibinfo {volume} {18}},\ \href
  {https://doi.org/10.1038/nn.4125} {10.1038/nn.4125} (\bibinfo {year}
  {2015})\BibitemShut {NoStop}%
\bibitem [{\citenamefont {Rodriguez}\ \emph {et~al.}(2019)\citenamefont
  {Rodriguez}, \citenamefont {Izquierdo},\ and\ \citenamefont
  {Ahn}}]{opt_mod_mcnr}%
  \BibitemOpen
  \bibfield  {author} {\bibinfo {author} {\bibfnamefont {N.}~\bibnamefont
  {Rodriguez}}, \bibinfo {author} {\bibfnamefont {E.}~\bibnamefont
  {Izquierdo}},\ and\ \bibinfo {author} {\bibfnamefont {Y.-Y.}\ \bibnamefont
  {Ahn}},\ }\bibfield  {title} {\bibinfo {title} {Optimal modularity and memory
  capacity of neural reservoirs},\ }\href
  {https://doi.org/10.1162/netn\_a\_00082} {\bibfield  {journal} {\bibinfo
  {journal} {Network Neuroscience}\ }\textbf {\bibinfo {volume} {3}},\ \bibinfo
  {pages} {551} (\bibinfo {year} {2019})}\BibitemShut {NoStop}%
\bibitem [{\citenamefont {Murphy}\ \emph {et~al.}(2020)\citenamefont {Murphy},
  \citenamefont {Bertolero}, \citenamefont {Papadopoulos}, \citenamefont
  {Lydon-Staley},\ and\ \citenamefont {Bassett}}]{murphy_2020}%
  \BibitemOpen
  \bibfield  {author} {\bibinfo {author} {\bibfnamefont {A.~C.}\ \bibnamefont
  {Murphy}}, \bibinfo {author} {\bibfnamefont {M.~A.}\ \bibnamefont
  {Bertolero}}, \bibinfo {author} {\bibfnamefont {L.}~\bibnamefont
  {Papadopoulos}}, \bibinfo {author} {\bibfnamefont {D.~M.}\ \bibnamefont
  {Lydon-Staley}},\ and\ \bibinfo {author} {\bibfnamefont {D.~S.}\ \bibnamefont
  {Bassett}},\ }\bibfield  {title} {\bibinfo {title} {Multimodal network
  dynamics underpinning working memory},\ }\bibfield  {journal} {\bibinfo
  {journal} {Nature Communications}\ }\textbf {\bibinfo {volume} {11}},\ \href
  {https://doi.org/10.1038/s41467-020-15541-0} {10.1038/s41467-020-15541-0}
  (\bibinfo {year} {2020})\BibitemShut {NoStop}%
\bibitem [{\citenamefont {Bassett}\ \emph {et~al.}(2018)\citenamefont
  {Bassett}, \citenamefont {Zurn},\ and\ \citenamefont {Gold}}]{numnn}%
  \BibitemOpen
  \bibfield  {author} {\bibinfo {author} {\bibfnamefont {D.}~\bibnamefont
  {Bassett}}, \bibinfo {author} {\bibfnamefont {P.}~\bibnamefont {Zurn}},\ and\
  \bibinfo {author} {\bibfnamefont {J.}~\bibnamefont {Gold}},\ }\bibfield
  {title} {\bibinfo {title} {On the nature and use of models in network
  neuroscience},\ }\bibfield  {journal} {\bibinfo  {journal} {Nature Reviews
  Neuroscience}\ }\textbf {\bibinfo {volume} {19}},\ \href
  {https://doi.org/10.1038/s41583-018-0038-8} {10.1038/s41583-018-0038-8}
  (\bibinfo {year} {2018})\BibitemShut {NoStop}%
\bibitem [{\citenamefont {Fox}\ \emph {et~al.}(2005)\citenamefont {Fox},
  \citenamefont {Snyder}, \citenamefont {Vincent}, \citenamefont {Corbetta},
  \citenamefont {Van~Essen},\ and\ \citenamefont {Raichle}}]{Fox05}%
  \BibitemOpen
  \bibfield  {author} {\bibinfo {author} {\bibfnamefont {M.~D.}\ \bibnamefont
  {Fox}}, \bibinfo {author} {\bibfnamefont {A.~Z.}\ \bibnamefont {Snyder}},
  \bibinfo {author} {\bibfnamefont {J.~L.}\ \bibnamefont {Vincent}}, \bibinfo
  {author} {\bibfnamefont {M.}~\bibnamefont {Corbetta}}, \bibinfo {author}
  {\bibfnamefont {D.~C.}\ \bibnamefont {Van~Essen}},\ and\ \bibinfo {author}
  {\bibfnamefont {M.~E.}\ \bibnamefont {Raichle}},\ }\bibfield  {title}
  {\bibinfo {title} {The human brain is intrinsically organized into dynamic,
  anticorrelated functional networks},\ }\href
  {https://doi.org/10.1073/pnas.0504136102} {\bibfield  {journal} {\bibinfo
  {journal} {Proceedings of the National Academy of Sciences}\ }\textbf
  {\bibinfo {volume} {102}},\ \bibinfo {pages} {9673} (\bibinfo {year}
  {2005})}\BibitemShut {NoStop}%
\bibitem [{\citenamefont {Damoiseaux}\ \emph {et~al.}(2006)\citenamefont
  {Damoiseaux}, \citenamefont {Rombouts}, \citenamefont {Barkhof},
  \citenamefont {Scheltens}, \citenamefont {Stam}, \citenamefont {Smith},\ and\
  \citenamefont {Beckmann}}]{Damoiseaux06}%
  \BibitemOpen
  \bibfield  {author} {\bibinfo {author} {\bibfnamefont {J.~S.}\ \bibnamefont
  {Damoiseaux}}, \bibinfo {author} {\bibfnamefont {S.~A. R.~B.}\ \bibnamefont
  {Rombouts}}, \bibinfo {author} {\bibfnamefont {F.}~\bibnamefont {Barkhof}},
  \bibinfo {author} {\bibfnamefont {P.}~\bibnamefont {Scheltens}}, \bibinfo
  {author} {\bibfnamefont {C.~J.}\ \bibnamefont {Stam}}, \bibinfo {author}
  {\bibfnamefont {S.~M.}\ \bibnamefont {Smith}},\ and\ \bibinfo {author}
  {\bibfnamefont {C.~F.}\ \bibnamefont {Beckmann}},\ }\bibfield  {title}
  {\bibinfo {title} {Consistent resting-state networks across healthy
  subjects},\ }\href {https://doi.org/10.1073/pnas.0601417103} {\bibfield
  {journal} {\bibinfo  {journal} {Proceedings of the National Academy of
  Sciences}\ }\textbf {\bibinfo {volume} {103}},\ \bibinfo {pages} {13848}
  (\bibinfo {year} {2006})}\BibitemShut {NoStop}%
\bibitem [{\citenamefont {Silverman}(1998)}]{kde_bandwidth}%
  \BibitemOpen
  \bibfield  {author} {\bibinfo {author} {\bibfnamefont {B.}~\bibnamefont
  {Silverman}},\ }\href@noop {} {\emph {\bibinfo {title} {Density estimation
  for statistics and data analysis}}}\ (\bibinfo  {publisher} {Chapman \&
  Hall/CRC},\ \bibinfo {address} {London},\ \bibinfo {year} {1998})\BibitemShut
  {NoStop}%
\bibitem [{\citenamefont {Soares}\ \emph {et~al.}(2016)\citenamefont {Soares},
  \citenamefont {Magalhães}, \citenamefont {Moreira}, \citenamefont {Sousa},
  \citenamefont {Ganz}, \citenamefont {Sampaio}, \citenamefont {Alves},
  \citenamefont {Marques},\ and\ \citenamefont {Sousa}}]{hitchhiker_fmri}%
  \BibitemOpen
  \bibfield  {author} {\bibinfo {author} {\bibfnamefont {J.~M.}\ \bibnamefont
  {Soares}}, \bibinfo {author} {\bibfnamefont {R.}~\bibnamefont {Magalhães}},
  \bibinfo {author} {\bibfnamefont {P.~S.}\ \bibnamefont {Moreira}}, \bibinfo
  {author} {\bibfnamefont {A.}~\bibnamefont {Sousa}}, \bibinfo {author}
  {\bibfnamefont {E.}~\bibnamefont {Ganz}}, \bibinfo {author} {\bibfnamefont
  {A.}~\bibnamefont {Sampaio}}, \bibinfo {author} {\bibfnamefont
  {V.}~\bibnamefont {Alves}}, \bibinfo {author} {\bibfnamefont
  {P.}~\bibnamefont {Marques}},\ and\ \bibinfo {author} {\bibfnamefont
  {N.}~\bibnamefont {Sousa}},\ }\bibfield  {title} {\bibinfo {title} {A
  hitchhiker's guide to functional magnetic resonance imaging},\ }\href
  {https://doi.org/10.3389/fnins.2016.00515} {\bibfield  {journal} {\bibinfo
  {journal} {Frontiers in Neuroscience}\ }\textbf {\bibinfo {volume} {10}},\
  \bibinfo {pages} {515} (\bibinfo {year} {2016})}\BibitemShut {NoStop}%
\bibitem [{\citenamefont {Armstrong}(2014)}]{bonferroni_review}%
  \BibitemOpen
  \bibfield  {author} {\bibinfo {author} {\bibfnamefont {R.~A.}\ \bibnamefont
  {Armstrong}},\ }\bibfield  {title} {\bibinfo {title} {When to use the
  bonferroni correction},\ }\href
  {https://doi.org/https://doi.org/10.1111/opo.12131} {\bibfield  {journal}
  {\bibinfo  {journal} {Ophthalmic and Physiological Optics}\ }\textbf
  {\bibinfo {volume} {34}},\ \bibinfo {pages} {502} (\bibinfo {year}
  {2014})}\BibitemShut {NoStop}%
\bibitem [{\citenamefont {Virtanen}\ \emph {et~al.}(2020)\citenamefont
  {Virtanen}, \citenamefont {Gommers}, \citenamefont {Oliphant}, \citenamefont
  {Haberland}, \citenamefont {Reddy}, \citenamefont {Cournapeau}, \citenamefont
  {Burovski}, \citenamefont {Peterson}, \citenamefont {Weckesser},
  \citenamefont {Bright}, \citenamefont {{van der Walt}}, \citenamefont
  {Brett}, \citenamefont {Wilson}, \citenamefont {Millman}, \citenamefont
  {Mayorov}, \citenamefont {Nelson}, \citenamefont {Jones}, \citenamefont
  {Kern}, \citenamefont {Larson}, \citenamefont {Carey}, \citenamefont {Polat},
  \citenamefont {Feng}, \citenamefont {Moore}, \citenamefont {{VanderPlas}},
  \citenamefont {Laxalde}, \citenamefont {Perktold}, \citenamefont {Cimrman},
  \citenamefont {Henriksen}, \citenamefont {Quintero}, \citenamefont {Harris},
  \citenamefont {Archibald}, \citenamefont {Ribeiro}, \citenamefont
  {Pedregosa}, \citenamefont {{van Mulbregt}},\ and\ \citenamefont {{SciPy 1.0
  Contributors}}}]{scipy}%
  \BibitemOpen
  \bibfield  {author} {\bibinfo {author} {\bibfnamefont {P.}~\bibnamefont
  {Virtanen}}, \bibinfo {author} {\bibfnamefont {R.}~\bibnamefont {Gommers}},
  \bibinfo {author} {\bibfnamefont {T.~E.}\ \bibnamefont {Oliphant}}, \bibinfo
  {author} {\bibfnamefont {M.}~\bibnamefont {Haberland}}, \bibinfo {author}
  {\bibfnamefont {T.}~\bibnamefont {Reddy}}, \bibinfo {author} {\bibfnamefont
  {D.}~\bibnamefont {Cournapeau}}, \bibinfo {author} {\bibfnamefont
  {E.}~\bibnamefont {Burovski}}, \bibinfo {author} {\bibfnamefont
  {P.}~\bibnamefont {Peterson}}, \bibinfo {author} {\bibfnamefont
  {W.}~\bibnamefont {Weckesser}}, \bibinfo {author} {\bibfnamefont
  {J.}~\bibnamefont {Bright}}, \bibinfo {author} {\bibfnamefont {S.~J.}\
  \bibnamefont {{van der Walt}}}, \bibinfo {author} {\bibfnamefont
  {M.}~\bibnamefont {Brett}}, \bibinfo {author} {\bibfnamefont
  {J.}~\bibnamefont {Wilson}}, \bibinfo {author} {\bibfnamefont {K.~J.}\
  \bibnamefont {Millman}}, \bibinfo {author} {\bibfnamefont {N.}~\bibnamefont
  {Mayorov}}, \bibinfo {author} {\bibfnamefont {A.~R.~J.}\ \bibnamefont
  {Nelson}}, \bibinfo {author} {\bibfnamefont {E.}~\bibnamefont {Jones}},
  \bibinfo {author} {\bibfnamefont {R.}~\bibnamefont {Kern}}, \bibinfo {author}
  {\bibfnamefont {E.}~\bibnamefont {Larson}}, \bibinfo {author} {\bibfnamefont
  {C.~J.}\ \bibnamefont {Carey}}, \bibinfo {author} {\bibfnamefont
  {{\.I}.}~\bibnamefont {Polat}}, \bibinfo {author} {\bibfnamefont
  {Y.}~\bibnamefont {Feng}}, \bibinfo {author} {\bibfnamefont {E.~W.}\
  \bibnamefont {Moore}}, \bibinfo {author} {\bibfnamefont {J.}~\bibnamefont
  {{VanderPlas}}}, \bibinfo {author} {\bibfnamefont {D.}~\bibnamefont
  {Laxalde}}, \bibinfo {author} {\bibfnamefont {J.}~\bibnamefont {Perktold}},
  \bibinfo {author} {\bibfnamefont {R.}~\bibnamefont {Cimrman}}, \bibinfo
  {author} {\bibfnamefont {I.}~\bibnamefont {Henriksen}}, \bibinfo {author}
  {\bibfnamefont {E.~A.}\ \bibnamefont {Quintero}}, \bibinfo {author}
  {\bibfnamefont {C.~R.}\ \bibnamefont {Harris}}, \bibinfo {author}
  {\bibfnamefont {A.~M.}\ \bibnamefont {Archibald}}, \bibinfo {author}
  {\bibfnamefont {A.~H.}\ \bibnamefont {Ribeiro}}, \bibinfo {author}
  {\bibfnamefont {F.}~\bibnamefont {Pedregosa}}, \bibinfo {author}
  {\bibfnamefont {P.}~\bibnamefont {{van Mulbregt}}},\ and\ \bibinfo {author}
  {\bibnamefont {{SciPy 1.0 Contributors}}},\ }\bibfield  {title} {\bibinfo
  {title} {{{SciPy} 1.0: Fundamental Algorithms for Scientific Computing in
  Python}},\ }\href {https://doi.org/10.1038/s41592-019-0686-2} {\bibfield
  {journal} {\bibinfo  {journal} {Nature Methods}\ }\textbf {\bibinfo {volume}
  {17}},\ \bibinfo {pages} {261} (\bibinfo {year} {2020})}\BibitemShut
  {NoStop}%
\bibitem [{\citenamefont {Seabold}\ and\ \citenamefont
  {Perktold}(2010)}]{seabold2010statsmodels}%
  \BibitemOpen
  \bibfield  {author} {\bibinfo {author} {\bibfnamefont {S.}~\bibnamefont
  {Seabold}}\ and\ \bibinfo {author} {\bibfnamefont {J.}~\bibnamefont
  {Perktold}},\ }\bibfield  {title} {\bibinfo {title} {statsmodels: Econometric
  and statistical modeling with python},\ }in\ \href@noop {} {\emph {\bibinfo
  {booktitle} {9th Python in Science Conference}}}\ (\bibinfo {year}
  {2010})\BibitemShut {NoStop}%
\bibitem [{\citenamefont {Mitchell}\ \emph {et~al.}(2013)\citenamefont
  {Mitchell}, \citenamefont {Lange},\ and\ \citenamefont
  {Brus}}]{mitchell2013gendered}%
  \BibitemOpen
  \bibfield  {author} {\bibinfo {author} {\bibfnamefont {S.~M.}\ \bibnamefont
  {Mitchell}}, \bibinfo {author} {\bibfnamefont {S.}~\bibnamefont {Lange}},\
  and\ \bibinfo {author} {\bibfnamefont {H.}~\bibnamefont {Brus}},\ }\bibfield
  {title} {\bibinfo {title} {Gendered citation patterns in international
  relations journals},\ }\href@noop {} {\bibfield  {journal} {\bibinfo
  {journal} {International Studies Perspectives}\ }\textbf {\bibinfo {volume}
  {14}},\ \bibinfo {pages} {485} (\bibinfo {year} {2013})}\BibitemShut
  {NoStop}%
\bibitem [{\citenamefont {Dion}\ \emph {et~al.}(2018)\citenamefont {Dion},
  \citenamefont {Sumner},\ and\ \citenamefont {Mitchell}}]{dion2018gendered}%
  \BibitemOpen
  \bibfield  {author} {\bibinfo {author} {\bibfnamefont {M.~L.}\ \bibnamefont
  {Dion}}, \bibinfo {author} {\bibfnamefont {J.~L.}\ \bibnamefont {Sumner}},\
  and\ \bibinfo {author} {\bibfnamefont {S.~M.}\ \bibnamefont {Mitchell}},\
  }\bibfield  {title} {\bibinfo {title} {Gendered citation patterns across
  political science and social science methodology fields},\ }\href@noop {}
  {\bibfield  {journal} {\bibinfo  {journal} {Political Analysis}\ }\textbf
  {\bibinfo {volume} {26}},\ \bibinfo {pages} {312} (\bibinfo {year}
  {2018})}\BibitemShut {NoStop}%
\bibitem [{\citenamefont {Caplar}\ \emph {et~al.}(2017)\citenamefont {Caplar},
  \citenamefont {Tacchella},\ and\ \citenamefont
  {Birrer}}]{caplar2017quantitative}%
  \BibitemOpen
  \bibfield  {author} {\bibinfo {author} {\bibfnamefont {N.}~\bibnamefont
  {Caplar}}, \bibinfo {author} {\bibfnamefont {S.}~\bibnamefont {Tacchella}},\
  and\ \bibinfo {author} {\bibfnamefont {S.}~\bibnamefont {Birrer}},\
  }\bibfield  {title} {\bibinfo {title} {Quantitative evaluation of gender bias
  in astronomical publications from citation counts},\ }\href@noop {}
  {\bibfield  {journal} {\bibinfo  {journal} {Nature Astronomy}\ }\textbf
  {\bibinfo {volume} {1}},\ \bibinfo {pages} {0141} (\bibinfo {year}
  {2017})}\BibitemShut {NoStop}%
\bibitem [{\citenamefont {Maliniak}\ \emph {et~al.}(2013)\citenamefont
  {Maliniak}, \citenamefont {Powers},\ and\ \citenamefont
  {Walter}}]{maliniak2013gender}%
  \BibitemOpen
  \bibfield  {author} {\bibinfo {author} {\bibfnamefont {D.}~\bibnamefont
  {Maliniak}}, \bibinfo {author} {\bibfnamefont {R.}~\bibnamefont {Powers}},\
  and\ \bibinfo {author} {\bibfnamefont {B.~F.}\ \bibnamefont {Walter}},\
  }\bibfield  {title} {\bibinfo {title} {The gender citation gap in
  international relations},\ }\href@noop {} {\bibfield  {journal} {\bibinfo
  {journal} {International Organization}\ }\textbf {\bibinfo {volume} {67}},\
  \bibinfo {pages} {889} (\bibinfo {year} {2013})}\BibitemShut {NoStop}%
\bibitem [{\citenamefont {Dworkin}\ \emph {et~al.}(2020)\citenamefont
  {Dworkin}, \citenamefont {Linn}, \citenamefont {Teich}, \citenamefont {Zurn},
  \citenamefont {Shinohara},\ and\ \citenamefont
  {Bassett}}]{Dworkin20200103894378}%
  \BibitemOpen
  \bibfield  {author} {\bibinfo {author} {\bibfnamefont {J.~D.}\ \bibnamefont
  {Dworkin}}, \bibinfo {author} {\bibfnamefont {K.~A.}\ \bibnamefont {Linn}},
  \bibinfo {author} {\bibfnamefont {E.~G.}\ \bibnamefont {Teich}}, \bibinfo
  {author} {\bibfnamefont {P.}~\bibnamefont {Zurn}}, \bibinfo {author}
  {\bibfnamefont {R.~T.}\ \bibnamefont {Shinohara}},\ and\ \bibinfo {author}
  {\bibfnamefont {D.~S.}\ \bibnamefont {Bassett}},\ }\bibfield  {title}
  {\bibinfo {title} {The extent and drivers of gender imbalance in neuroscience
  reference lists},\ }\bibfield  {journal} {\bibinfo  {journal} {bioRxiv}\
  }\href {https://doi.org/10.1101/2020.01.03.894378}
  {10.1101/2020.01.03.894378} (\bibinfo {year} {2020})\BibitemShut {NoStop}%
\bibitem [{\citenamefont {Bertolero}\ \emph {et~al.}(2020)\citenamefont
  {Bertolero}, \citenamefont {Dworkin}, \citenamefont {David}, \citenamefont
  {Lloreda}, \citenamefont {Srivastava}, \citenamefont {Stiso}, \citenamefont
  {Zhou}, \citenamefont {Dzirasa}, \citenamefont {Fair}, \citenamefont
  {Kaczkurkin}, \citenamefont {Marlin}, \citenamefont {Shohamy}, \citenamefont
  {Uddin}, \citenamefont {Zurn},\ and\ \citenamefont
  {Bassett}}]{bertolero2021racial}%
  \BibitemOpen
  \bibfield  {author} {\bibinfo {author} {\bibfnamefont {M.~A.}\ \bibnamefont
  {Bertolero}}, \bibinfo {author} {\bibfnamefont {J.~D.}\ \bibnamefont
  {Dworkin}}, \bibinfo {author} {\bibfnamefont {S.~U.}\ \bibnamefont {David}},
  \bibinfo {author} {\bibfnamefont {C.~L.}\ \bibnamefont {Lloreda}}, \bibinfo
  {author} {\bibfnamefont {P.}~\bibnamefont {Srivastava}}, \bibinfo {author}
  {\bibfnamefont {J.}~\bibnamefont {Stiso}}, \bibinfo {author} {\bibfnamefont
  {D.}~\bibnamefont {Zhou}}, \bibinfo {author} {\bibfnamefont {K.}~\bibnamefont
  {Dzirasa}}, \bibinfo {author} {\bibfnamefont {D.~A.}\ \bibnamefont {Fair}},
  \bibinfo {author} {\bibfnamefont {A.~N.}\ \bibnamefont {Kaczkurkin}},
  \bibinfo {author} {\bibfnamefont {B.~J.}\ \bibnamefont {Marlin}}, \bibinfo
  {author} {\bibfnamefont {D.}~\bibnamefont {Shohamy}}, \bibinfo {author}
  {\bibfnamefont {L.~Q.}\ \bibnamefont {Uddin}}, \bibinfo {author}
  {\bibfnamefont {P.}~\bibnamefont {Zurn}},\ and\ \bibinfo {author}
  {\bibfnamefont {D.~S.}\ \bibnamefont {Bassett}},\ }\bibfield  {title}
  {\bibinfo {title} {Racial and ethnic imbalance in neuroscience reference
  lists and intersections with gender},\ }\href@noop {} {\bibfield  {journal}
  {\bibinfo  {journal} {bioRxiv}\ } (\bibinfo {year} {2020})}\BibitemShut
  {NoStop}%
\bibitem [{\citenamefont {Wang}\ \emph {et~al.}(2021)\citenamefont {Wang},
  \citenamefont {Dworkin}, \citenamefont {Zhou}, \citenamefont {Stiso},
  \citenamefont {Falk}, \citenamefont {Bassett}, \citenamefont {Zurn},\ and\
  \citenamefont {Lydon-Staley}}]{wang2021gendered}%
  \BibitemOpen
  \bibfield  {author} {\bibinfo {author} {\bibfnamefont {X.}~\bibnamefont
  {Wang}}, \bibinfo {author} {\bibfnamefont {J.~D.}\ \bibnamefont {Dworkin}},
  \bibinfo {author} {\bibfnamefont {D.}~\bibnamefont {Zhou}}, \bibinfo {author}
  {\bibfnamefont {J.}~\bibnamefont {Stiso}}, \bibinfo {author} {\bibfnamefont
  {E.~B.}\ \bibnamefont {Falk}}, \bibinfo {author} {\bibfnamefont {D.~S.}\
  \bibnamefont {Bassett}}, \bibinfo {author} {\bibfnamefont {P.}~\bibnamefont
  {Zurn}},\ and\ \bibinfo {author} {\bibfnamefont {D.~M.}\ \bibnamefont
  {Lydon-Staley}},\ }\bibfield  {title} {\bibinfo {title} {Gendered citation
  practices in the field of communication},\ }\bibfield  {journal} {\bibinfo
  {journal} {Annals of the International Communication Association}\ }\href
  {https://doi.org/10.1080/23808985.2021.1960180}
  {10.1080/23808985.2021.1960180} (\bibinfo {year} {2021})\BibitemShut
  {NoStop}%
\bibitem [{\citenamefont {Chatterjee}\ and\ \citenamefont
  {Werner}(2021)}]{chatterjee2021gender}%
  \BibitemOpen
  \bibfield  {author} {\bibinfo {author} {\bibfnamefont {P.}~\bibnamefont
  {Chatterjee}}\ and\ \bibinfo {author} {\bibfnamefont {R.~M.}\ \bibnamefont
  {Werner}},\ }\bibfield  {title} {\bibinfo {title} {Gender disparity in
  citations in high-impact journal articles},\ }\href@noop {} {\bibfield
  {journal} {\bibinfo  {journal} {JAMA Netw Open}\ }\textbf {\bibinfo {volume}
  {4}},\ \bibinfo {pages} {e2114509} (\bibinfo {year} {2021})}\BibitemShut
  {NoStop}%
\bibitem [{\citenamefont {Fulvio}\ \emph {et~al.}(2021)\citenamefont {Fulvio},
  \citenamefont {Akinnola},\ and\ \citenamefont
  {Postle}}]{fulvio2021imbalance}%
  \BibitemOpen
  \bibfield  {author} {\bibinfo {author} {\bibfnamefont {J.~M.}\ \bibnamefont
  {Fulvio}}, \bibinfo {author} {\bibfnamefont {I.}~\bibnamefont {Akinnola}},\
  and\ \bibinfo {author} {\bibfnamefont {B.~R.}\ \bibnamefont {Postle}},\
  }\bibfield  {title} {\bibinfo {title} {Gender (im)balance in citation
  practices in cognitive neuroscience},\ }\href@noop {} {\bibfield  {journal}
  {\bibinfo  {journal} {J Cogn Neurosci}\ }\textbf {\bibinfo {volume} {33}},\
  \bibinfo {pages} {3} (\bibinfo {year} {2021})}\BibitemShut {NoStop}%
\bibitem [{\citenamefont {Zhou}\ \emph
  {et~al.}(2020{\natexlab{b}})\citenamefont {Zhou}, \citenamefont {Cornblath},
  \citenamefont {Stiso}, \citenamefont {Teich}, \citenamefont {Dworkin},
  \citenamefont {Blevins},\ and\ \citenamefont
  {Bassett}}]{zhou_dale_2020_3672110}%
  \BibitemOpen
  \bibfield  {author} {\bibinfo {author} {\bibfnamefont {D.}~\bibnamefont
  {Zhou}}, \bibinfo {author} {\bibfnamefont {E.~J.}\ \bibnamefont {Cornblath}},
  \bibinfo {author} {\bibfnamefont {J.}~\bibnamefont {Stiso}}, \bibinfo
  {author} {\bibfnamefont {E.~G.}\ \bibnamefont {Teich}}, \bibinfo {author}
  {\bibfnamefont {J.~D.}\ \bibnamefont {Dworkin}}, \bibinfo {author}
  {\bibfnamefont {A.~S.}\ \bibnamefont {Blevins}},\ and\ \bibinfo {author}
  {\bibfnamefont {D.~S.}\ \bibnamefont {Bassett}},\ }\href
  {https://doi.org/10.5281/zenodo.3672110} {\bibinfo {title} {Gender diversity
  statement and code notebook v1.0}} (\bibinfo {year}
  {2020}{\natexlab{b}})\BibitemShut {NoStop}%
\bibitem [{\citenamefont {Ambekar}\ \emph {et~al.}(2009)\citenamefont
  {Ambekar}, \citenamefont {Ward}, \citenamefont {Mohammed}, \citenamefont
  {Male},\ and\ \citenamefont {Skiena}}]{ambekar2009name}%
  \BibitemOpen
  \bibfield  {author} {\bibinfo {author} {\bibfnamefont {A.}~\bibnamefont
  {Ambekar}}, \bibinfo {author} {\bibfnamefont {C.}~\bibnamefont {Ward}},
  \bibinfo {author} {\bibfnamefont {J.}~\bibnamefont {Mohammed}}, \bibinfo
  {author} {\bibfnamefont {S.}~\bibnamefont {Male}},\ and\ \bibinfo {author}
  {\bibfnamefont {S.}~\bibnamefont {Skiena}},\ }\bibfield  {title} {\bibinfo
  {title} {Name-ethnicity classification from open sources},\ }in\ \href@noop
  {} {\emph {\bibinfo {booktitle} {Proceedings of the 15th ACM SIGKDD
  international conference on Knowledge Discovery and Data Mining}}}\ (\bibinfo
  {year} {2009})\ pp.\ \bibinfo {pages} {49--58}\BibitemShut {NoStop}%
\bibitem [{\citenamefont {Sood}\ and\ \citenamefont
  {Laohaprapanon}(2018)}]{sood2018predicting}%
  \BibitemOpen
  \bibfield  {author} {\bibinfo {author} {\bibfnamefont {G.}~\bibnamefont
  {Sood}}\ and\ \bibinfo {author} {\bibfnamefont {S.}~\bibnamefont
  {Laohaprapanon}},\ }\bibfield  {title} {\bibinfo {title} {Predicting race and
  ethnicity from the sequence of characters in a name},\ }\href@noop {}
  {\bibfield  {journal} {\bibinfo  {journal} {arXiv preprint arXiv:1805.02109}\
  } (\bibinfo {year} {2018})}\BibitemShut {NoStop}%
\bibitem [{\citenamefont {{Van Essen}}\ \emph {et~al.}(2013)\citenamefont {{Van
  Essen}}, \citenamefont {Smith}, \citenamefont {Barch}, \citenamefont
  {Behrens}, \citenamefont {Yacoub},\ and\ \citenamefont
  {Ugurbil}}]{hcp_overview}%
  \BibitemOpen
  \bibfield  {author} {\bibinfo {author} {\bibfnamefont {D.~C.}\ \bibnamefont
  {{Van Essen}}}, \bibinfo {author} {\bibfnamefont {S.~M.}\ \bibnamefont
  {Smith}}, \bibinfo {author} {\bibfnamefont {D.~M.}\ \bibnamefont {Barch}},
  \bibinfo {author} {\bibfnamefont {T.~E.}\ \bibnamefont {Behrens}}, \bibinfo
  {author} {\bibfnamefont {E.}~\bibnamefont {Yacoub}},\ and\ \bibinfo {author}
  {\bibfnamefont {K.}~\bibnamefont {Ugurbil}},\ }\bibfield  {title} {\bibinfo
  {title} {The wu-minn human connectome project: An overview},\ }\href
  {https://doi.org/https://doi.org/10.1016/j.neuroimage.2013.05.041} {\bibfield
   {journal} {\bibinfo  {journal} {NeuroImage}\ }\textbf {\bibinfo {volume}
  {80}},\ \bibinfo {pages} {62} (\bibinfo {year} {2013})},\ \bibinfo {note}
  {mapping the Connectome}\BibitemShut {NoStop}%
\end{thebibliography}%


%

\clearpage
\appendix
\section{Euclidean distance as a covariate}

\subsection{Relationship with Minimum Control Energy}
\label{sec:app_ed_mce}

\noindent
In the present experiments, the control set was set to the identity matrix. 
The expression of the controllability Gramian thus simplifies considerably to
\begin{equation}
    \mathbf{W_C} = \int_0^T \mathrm{e}^{\mathbf{A}t} \mathrm{e}^{\mathbf{A}^\top t} dt.
\end{equation}
As a diffusion connectivity matrix, $\mathbf{A}$ is symmetric, such that $\mathbf{A}=\mathbf{A}^\top$, thus the expression can be evaluated analytically:
\begin{equation}
    \mathbf{W_C} = \int_0^T \mathrm{e}^{2\mathbf{A}t} = \frac{\mathrm{e}^{2\mathbf{A}T}-\mathbf{I}}{2\mathbf{A}}
\end{equation}
Thus, its inverse is
\begin{equation}
    \mathbf{W_C}^{-1} = \frac{2\mathbf{A}}{\mathrm{e}^{2\mathbf{A}T}-\mathbf{I}}.
\end{equation}
Taking the Taylor series expansion of this equation about $\mathbf{A}=0$ yields
\begin{equation}
    \mathbf{W_C}^{-1} = \frac{\mathbf{I}}{T} - \mathbf{A} + \frac{1}{3}\mathbf{A}^2 T + \mathcal{O}(\mathbf{A}^4).
\end{equation}
The minimum control energy $E_{min}$ (see Eq.~\ref{eq:e_min}) can be written as
\begin{equation}
    E_{min} = \mathbf{x}^\top \mathbf{W_C}^{-1} \mathbf{x},
\end{equation}
with $\mathbf{x}$ the current brain state.
Setting $T=10$, as in all experiments, the first-order approximation of the minimum control energy thus becomes
\begin{equation}
    E_{min} \approx \frac{1}{10}\mathbf{x}^\top \mathbf{x} - \mathbf{x}^\top \mathbf{A} \mathbf{x}.
\end{equation}

In sum, the minimum energy is approximately equal to the weighted sum of magnitude of the squared Euclidean distance of the state, $\mathbf{x}^\top \mathbf{x}$ and the projection of that distance onto $\mathbf{A}$.
It should be noted that $\mathbf{A}$ designates the post-normalized matrix.
This normalization makes all eigenvalues of $\mathbf{A}$ negative, such that the contributing term $ - \mathbf{x}^\top \mathbf{A} \mathbf{x}$ is always positive.

\subsection{Relationship with Information Content}
\label{sec:app_ed_ic}

\noindent
The brain activations in fMRI experiments are unimodal, and all timeseries were demeaned.
Under the assumption of 1) Gaussian distribution of fMRI time series and 2) equal standard deviation across brain participants, the fMRI times series of single parcels would follow \mbox{$X\sim\mathcal{N}(0,\sigma^2)$} with:
\begin{equation}
    \mathcal{N}(0,\sigma^2) = \frac{1}{\sigma \sqrt{2 \pi}} \mathrm{e}^{-0.5(\frac{x_{\rm{parcel}}}{\sigma})^2}.
\end{equation}
As information content is defined based on the PDF with $I = -\log(X)$, the information content of a single parcel is
\begin{equation}
    I_{\rm{parcel}} = \log (\sigma \sqrt{2 \pi} ) + 0.5 (\frac{x_{\rm{parcel}}}{\sigma})^2.
\end{equation}
For the given assumptions, $I_{\rm{parcel}}$ thus follows the squared activation intensity, with an offset depending on the standard deviation of the distribution.
For whole brain information content, $I_{\rm{parcel}}$ is summed over all parcels:
\begin{equation}
    I_{\rm{state}} = \sum_{\rm{parcels}} I_{\rm{parcel}} = n\log (\sigma \sqrt{2 \pi} ) + \frac{n}{2\sigma^2} \mathbf{x}^\top \mathbf{x},
\end{equation}
$n$ being the number of parcels.
This results in a direct correlation to squared Euclidean distance of the state.

In reality, fMRI activations are not completely Gaussian, and the amplitude within- and between subjects varies~\cite{fMRI_amp_fluct}.
However, a correlation between squared Euclidean distance and information content was still observed.
The Pearson correlation coefficient (across all states, all tasks and all subjects) between squared Euclidean distance and $I_{\rm{state}}$ was 0.54, between squared Euclidean distance and $E_{\rm{min}}$ 0.65; and between $I_{\rm{state}}$ and $E_{\rm{min}}$ 0.91.

\end{document}